\documentclass[11pt]{article}
\usepackage{graphicx,amsmath,amsfonts,amscd,amssymb,bm,cite,epsfig,epsf,url,color,algorithm,algorithmic,multirow}
\usepackage{fullpage} \usepackage[small,bf]{caption}
\usepackage{subfigure}
\numberwithin{equation}{section}

\newcommand{\Ss}{{\mathcal S}}
\newcommand{\R}{\mathbb{R}}

\title{Online Robust Subspace Tracking from Partial Information}

\author{Jun~He$^{\sharp}$, Laura~Balzano$^{\dagger}$, and John~C.S.~Lui$^{\ddagger}$\\
 \vspace{-.1cm}\\        
 $^{\sharp}$ College of Electronic and Information Engineering, Nanjing University of \\Information  Science and Technology, Nanjing, 210044, China\\ hejun.zz@gmail.com\\
  \vspace{-.3cm}\\
  $^{\dagger}$ Department of Electrical and Computer Engineering, \\University of Wisconsin-Madison, Madison, WI, 53706, USA\\
sunbeam@ece.wisc.edu \\
  \vspace{-.3cm}\\
  $^{\ddagger}$ Department of Computer Science and Engineering, \\The Chinese University of Hong Kong, N.T., Hong Kong\\
  cslui@cse.cuhk.edu.hk\\
  \vspace{-.3cm}\\
}
        
\date{September 17, 2011}

\begin{document}
%\thanks{Jun He is with the College of Electronic and Information Engineering, 
%Nanjing University of Information Science and Technology, Nanjing,
%Jiangsu, China (hejun.zz@gmail.com).}% <-this % stops a space
%\thanks{Laura Balzano is with the Department of Electrical and Computer Engineering, University of Wisconsin-Madison,
%Wisconsin, U.S., (sunbeam@ece.wisc.edu).}% <-this % stops a space
%\thanks{John C.S. Lui is with the Department of Computer Science and Engineering, the Chinese University of Hong Kong,
%Hong Kong, (cslui@cse.cuhk.edu.hk).}% <-this % stops a space
\maketitle 

\begin{abstract}
This paper presents GRASTA (Grassmannian Robust Adaptive Subspace Tracking Algorithm), an efficient and robust online algorithm for tracking subspaces from highly incomplete information. The algorithm uses a robust $l^1$-norm cost function in order to estimate and track non-stationary subspaces when the streaming data vectors are corrupted with outliers. We apply GRASTA to the problems of robust matrix completion and real-time separation of background from foreground in video. In this second application, we show that GRASTA performs high-quality separation of moving objects from background at exceptional speeds: In one popular benchmark video example~\cite{Li04}, GRASTA achieves a rate of 57 frames per second, even when run in MATLAB on a personal laptop.
%\boldmath
\end{abstract}


%\IEEEpeerreviewmaketitle

\section{Introduction}

%\IEEEPARstart{L}
Low-rank subspaces have long been a powerful tool in data modeling and analysis. Applications in communications~\cite{Moulines95}, source localization and target tracking in radar and sonar~\cite{KrimViberg}, and medical imaging~\cite{Audette2000} all leverage subspace models in order to recover the signal of interest and reject noise. In these classical signal processing problems, a handful of high-quality sensors are co-located such that data can be reliably collected. 

The challenges of modern data analysis breach this standard setup. A first difference, one that cannot be overstated, is that data are being collected everywhere, on a more massive scale than ever before, by cameras, sensors, and people. We give just a few examples: There are an estimated minimum 10,000 surveillance cameras in the city of Chicago and an estimated 500,000 in London~\cite{chicagocams10, CCTV2002}. Netflix collects ratings from 25 million users on tens of thousands of movies~\cite{netflixPR}. On its peak day of the holiday season in 2008, Amazon.com collected data on 72 items purchased every second~\cite{amazonSales}. The 
Large Synoptic Survey Telescope, which will be deployed in Chile and will photograph the whole sky visible to it every three nights, will produce 20 terabytes of data every night~\cite{LSST}.

A second and equally important difference is that, in all these examples mentioned, the data collected may be unreliable or an indirect indicator of what one really wants to know. The data are collected from many possibly distributed sensors or even from people whose responses may be inconsistent, and the data may be missing or corrupted.

In order to address these issues, algorithms for data analysis must be computationally fast as well as robust to corruption and missing data. In this paper we present the Grassmannian Robust Adaptive Subspace Tracking Algorithm, or GRASTA, an online algorithm for robust subspace tracking that handles these three challenges at once. We seek a low-rank model for data that may be corrupted by outliers and have missing data values.  

GRASTA uses the natural $l^1$-norm cost function for data corrupted by sparse outliers, and performs incremental gradient descent on the Grassmannian, the manifold of all $d$-dimensional subspaces for fixed $d$. For each subspace update, we use the gradient of the augmented Lagrangian
function associated to this cost. GRASTA operates only one data vector at a time, making it faster than other state-of-the-art algorithms and amenable to streaming and real-time applications.

\subsection{Contributions}
The contributions of our work are threefold:

\subsubsection{Efficient Grassmannian Augmented Lagrangian Alternating Direction Algorithm}
We propose an efficient online robust subspace tracking algorithm -- GRASTA, or Grassmannian Robust Adaptive Subspace Tracking Algorithm -- which combines the augmented Lagrangian function with the classic stochastic gradient framework~\cite{kushner2003stochastic} and the structure of the Grassmannian~\cite{Edelman98}, and solves via the augmented Lagrangian alternating direction method~\cite{boyd2010distributed}. As we discuss in detail in Section~\ref{sec:prob-set-up} and~\ref{sec:alg_derivation}, GRASTA alternates between estimating a low-dimensional subspace $\Ss$ and a triple $(s, w, y)$ which represent the sparse corruptions in the signal, the weights for the fit of the signal to the subspace, and the dual vector in the optimization problem. For estimating the subspace $\Ss$, GRASTA uses gradient descent on the Grassmannian with $(s,w,y)$ fixed; for estimating the triple $(s,w,y)$, GRASTA uses ADMM~\cite{boyd2010distributed}.

When data vectors arise from an underlying subspace which is inherently low-dimensional, and are corrupted with noise and outliers, GRASTA is able estimate and track the subspace successfully, even when the vectors are highly incomplete.

\subsubsection{Fast Robust Low-rank Matrix Completion}
We show that GRASTA can successfully recover a low-rank matrix from partial information, even if the partially observed entries are corrupted by gross outliers. GRASTA's incremental update results in a significant speed-up over other state-of-the-art robust matrix completion algorithms or RPCA (robust principal components analysis) algorithms.

\subsubsection{Realtime Separation of Background and Moving Objects in Video Surveillance }
Finally, we show that the online nature of GRASTA makes it suitable for realtime high-dimensional sparse signal separation from a background signal, such as the task of separating background and moving objects in video surveillance. Compared to other RPCA methods, GRASTA can handle video frames at very high rates-- up to 57 frames per second in our examples-- even when implemented in MATLAB on a personal laptop, which is a significant practical advantage over other state-of-the-art techniques.

\vspace{.2in}
This paper is organized as follows. We motivate robust online subspace tracking and give background on subspace tracking and matrix completion in Sections~\ref{sec:motivation} and~\ref{sec:background}. The familiar reader can go directly to Section~\ref{sec:prob-set-up}, where we formulate the robust subspace tracking problem and 
introduce the novel subspace error function in Section~\ref{sec:prob-set-up}. In Section~\ref{sec:alg_derivation}, we present the Grassmannian 
Robust Adaptive Subspace Tracking Algorithm (GRASTA) in detail and discuss critical parts of the implementation; we point out the limitations and merits as compared with other RPCA algorithms. In Section~\ref{sec:expts}, we compare GRASTA with GROUSE and RPCA algorithms via extensive numerical experiments and several real-world video surveillance experiments. Section~\ref{sec:conclusion} concludes our work and gives some discussion on future directions.

\subsection{Motivations}
\label{sec:motivation}
\subsubsection{Online Subspace Tracking within Outliers}

GRASTA is built on GROUSE~\cite{grouse}, an efficient online subspace tracking algorithm. GROUSE uses an $l^2$-norm cost function, which is problematic when facing data corruption or noise distributed other than Gaussian.

As an example, we consider using subspaces to detect anomalies in computer networks~\cite{lakhina04}. A non-robust subspace estimation algorithm like GROUSE would need a special anomaly detection component in order to differentiate anomalies and outliers from the underlying subspace of the traffic data. Often these types of anomaly detection components rely on a lot of parameter tuning and heuristic rules for detection. This motivates a more principled approach that is robust by design: GRASTA.

\subsubsection{Robust Principal Component Analysis }
Principal Components Analysis~\cite{JoliffePCA} is a critical tool for data analysis in many fields. Given a parameter $d$ for the number of components desired, PCA seeks to find the best-fit (in an $l^2$ norm sense) $d$-dimensional subspace to data; in other words, it finds the best $d$ vectors, the principal components, such that the data can be approximated by a linear combination of those $d$ vectors.

The residuals of an $l^2$-norm error function will be Gaussian distributed. Therefore, even with one outlier data point, the principal components can be arbitrarily far from those without the outlier data point~\cite{RobustStatistics}. Modern data applications-- such as those in sensor networks, collaborative filtering, video surveillance or the network monitoring example just given-- will all experience data failures that result in outliers. Sometimes the outliers are even the signal of interest, as in the case of network anomaly detection or identifying moving objects in the foreground of a surveillance camera. 

A good deal of research is therefore focused on Robust PCA, including~\cite{chandrasekaran2011rank,CandesRPCA09}. Recent work focuses on a problem definition which seeks a low-rank and sparse matrix whose sum is the observed data. The majority of algorithms use SVD (singular value decomposition) computations to perform Robust PCA. The SVD is too slow for many real-time applications, and consequently many online SVD and subspace identification algorithms have been developed, as we discuss in Section~\ref{sec:onlineSTbackground}. We are therefore motivated to bridge the gap between online algorithms and robust algorithms with GRASTA.

Of course we emphasize that besides the ability to do matrix separation into low-rank and sparse parts, GRASTA can also effectively handle the scenario where the low-rank subspace is dynamic.

\subsection{Background}
\label{sec:background}
%\IEEEpubidadjcol

\subsubsection{Subspace Tracking}

First we briefly describe the subspace tracking problem set-up and GROUSE~\cite{grouse} algorithm before reviewing previous literature on subspace tracking. 

Consider a sequence of $d$-dimensional subspaces $\Ss_t \subset \R^n$, $d<n$, and a sequence of vectors $v_t \in \Ss_t$. The object of a subspace tracking algorithm is to estimate $\Ss_t$, given only $v_t$ and the previous subspace estimate $\Ss_{t-1}$. %The tolerance depends on how quickly the subspace is changing, but heuristically, successful subspace tracking algorithms will do well if the subspace changes slowly enough.

\vspace{.1in}
\paragraph{\em{Incomplete Data Vectors}}

Now considering the issue of incomplete data vectors, the object of an algorithm for subspace tracking with missing data is to estimate $\Ss_t$ given $v_{\Omega_t}$-- an incomplete version of $v_t$, observed only on the indices $\Omega \subset \{1, \dots, n\}$. The GROUSE~\cite{grouse} algorithm addresses exactly this problem. GROUSE is an incremental gradient descent algorithm performed on the Grassmannian $\mathcal{G} (d,n)$, the space of all $d$-dimensional subspaces of $\R^n$. The algorithm minimizes an $l^2$-norm cost between observed incomplete vectors and their fit to the subspace variable. Each step of the algorithm is simple and requires very few operations. However, the use of the $l^2$ loss makes GROUSE very susceptible to outliers.

\vspace{.1in}
\paragraph{\em{Complete Data Vectors}} 
\label{sec:onlineSTbackground}
Comon and Golub~\cite{comon90} give an early survey of adaptive methods for tracking subspaces, both coming from the matrix computation literature, including Lanczos-based recursion algorithms, and gradient-based methods from the signal processing literature. 

There is a vast literature on the adaptation of QR and SVD factorizations to the adaptive, online context. The work in~\cite{bischof92,moonen92,stewart92} are all along these lines. The fastest algorithm for incremental SVD is given in~\cite{Brand06}; this algorithm makes modifications, one column at a time, to the thin SVD of a strictly rank-$d$ $n \times n$ matrix in $O(n^2d)$ time.

Initial work in signal processing for subspace tracking was aimed at estimating from data the largest eigensubspace for a signal covariance matrix. 
This is useful, for example, in direction-of-arrival (DOA) estimation: the well-known work in~\cite{Roy89} introduces ESPRIT, a parameter estimation algorithm that estimates the DOA of plane waves emanating from a target and being received by a sensor array. ESPRIT was a follow up to the MUSIC algorithm~\cite{schmidt86}, and ESPRIT gains computational efficiency over MUSIC for a slight tradeoff in generality of sensor array design. Around the same time, Yang and Kaveh~\cite{YangKaveh} introduced an approach for subspace tracking that, like GROUSE, uses incremental gradient, thus making it more suitable for adaptive estimation of the signal subspace and covariance matrix. This work was followed by~\cite{mathew95, yang95, DelmasCardoso} with various improvements and convergence analyses. Unlike GROUSE, these algorithms all conduct gradient descent in the ambient space as opposed to operating along the geodesics of the Grassmannian. Also unlike GROUSE, these algorithms all require fully observed vectors.

Smith~\cite{EdelmanSmith96, STSthesis, Edelman98} thoroughly pursued conjugate gradient descent methods on the Grassmannian for solving the subspace tracking problem using the Rayleigh quotient as a cost function as opposed to the Frobenius norm of GROUSE. In~\cite{EdelmanSmith96} the authors give a very careful definition of the problem, giving a nice survey comparing the applicability of various approaches. In~\cite{Edelman98} is an extensive list of subspace tracking references.

We note here that none of the work in this subsection addressed issues of robustness to corrupted data or missing data.

\vspace{.1in}
\paragraph{\em{Robust Subspace Tracking}}
The work of~\cite{mateos10} addresses the problem of robust online subspace tracking. They focus on the problem where outliers are found in a fraction of vectors (that is, some vectors have no outliers), though they do remark that this can be extended to handle the case where outliers are sparse in every vector. They have a very nice proposition relating $l^0$-(pseudo)norm minimization to the least trimmed squares estimator. 

We note here that GRASTA differs from~\cite{mateos10} in that it directly focuses on the case where every vector may have outliers, it operates on the Grassmannian for greater efficiency, and it can handle missing data. A comparison to~\cite{mateos10} is a subject of future investigation.

\subsubsection{Matrix Completion}

The popular Netflix prize~\cite{NetflixPrize} stimulated research on the matrix completion problem: Given very few entries of a low-rank matrix, can one recover (or complete) the entire matrix? When Cand\`{e}s and Recht proved that, under some incoherence conditions, nuclear norm minimization recovers a highly incomplete low rank matrix with high probability~\cite{CandesRecht09}, an entire area was opened up for further analysis and algorithmic variations. Algorithms that have been proposed to solve matrix completion include ADMiRA~\cite{Lee09}, OptSpace~\cite{Keshavan10b}, Singular Value Thresholding~\cite{Cai08}, FPCA~\cite{Ma11}, SET~\cite{dai11}, APGL~\cite{Toh10}, GROUSE~\cite{grouse}, and many others. Of these, GROUSE is the only {\em online} matrix completion algorithm in that it proceeds incrementally, one column at a time. This along with the fact that each update of GROUSE has low computational complexity makes GROUSE the fastest of the state-of-the-art matrix completion algorithms by nearly an order of magnitude~\cite{grouse}.

% Problem Setup

\section{Problem Set-up}
\label{sec:prob-set-up}

We denote the evolving $d$-dimensional subspace of $\mathbb{R}^n$ as $\Ss_t$ at time $t$. In applications of interest we have $d\ll n$. 
Let the columns of an $n\times d$ matrix $U_t$ be orthonormal and span $\mathcal{S}_t$. Tracking the evolving subspace $\mathcal{S}_t$ is equivalent to estimating $U_t$ at each time step\footnote{We remind the reader here that $U_t$ is not unique for a given subspace, but the projection matrix $U_t U_t^T$ is unique.}.

\subsection{Model}
At each time step $t$, we assume that $v_t$ is generated by the following  model: %\ref{eq:measmodel}:

\begin{equation}
	v_t = U_t w_t + s_t + \zeta_t
	\label{eq:measmodel}
\end{equation}
where $w_t$ is the $d \times 1$ weight vector, $s_t$ is the $n \times 1$ sparse outlier vector whose nonzero entries may be arbitrarily large, and $\zeta_t$ is the $n \times 1$ zero-mean Gaussian white noise vector with small variance. We observe only a small subset of entries of $v_t$, denoted by $\Omega_t \subset \{1, \dots, n\}$. 

Conforming to the notation of GROUSE ~\cite{grouse}, we let $U_{\Omega_t}$ denote the submatrix  of $U_t$ consisting of the rows indexed by $\Omega_t$; also for a vector $v_t \in \mathbb{R}^n$,  let $v_{\Omega_t}$% = \mathcal{P}_{\Omega_t}(v_t)$ 
denote a vector in $\mathbb{R}^{|\Omega_t|}$ whose entries are those of $v_t$ indexed by $\Omega_t$. A critical problem raised when we only partially observe $v_t$ is how to quantify the subspace error only from the incomplete and corrupted data. GROUSE~\cite{grouse} uses the natural Euclidean distance, the $l^2$-norm, to measure the subspace error from the subspace spanned by the columns of $U_t$ to the observed vector $v_{\Omega_t}$:  

\begin{equation}
F_{grouse}(\Ss;t) = \min_w \| U_{\Omega_t}w - v_{\Omega_t} \|_2^2\;.
\label{eq:grousecost}
\end{equation} It was shown in~\cite{Balzano10a} that this cost function gives an accurate estimate of the same cost function with full data ($\Omega=\{1,\dots,n\}$), as long as $|\Omega_t|$ is large enough\footnote{In~\cite{Balzano10a} the authors show that $|\Omega_t|$ must be larger than $\mu(\Ss) d \log(2d/\delta)$, where $\mu(\Ss)$ is a measure of incoherence on the subspace and $\delta$ controls the probability of the result. See the paper for details.}. However, if the observed data vector is corrupted by outliers as in Equation \eqref{eq:measmodel}, 
an $l^2$-based best-fit to the subspace can be influenced arbitrarily with just one large outlier; this in turn will lead to an incorrect subspace update in the GROUSE algorithm, as we demonstrate in Section \ref{sec:GROUSEcompare}.

\subsection{Subspace Error Quantification by $l^1$-Norm }
In order to quantify the subspace error robustly, we use the $l^1$-norm as follows: 

\begin{equation}\label{eq:L1_distance}
	F_{grasta}(\Ss; t) = \min_w{\|  U_{\Omega_t}w-v_{\Omega_t} \|_1}\;.
\end{equation}
With $U_{\Omega_t}$ known (or estimated, but fixed), this $l^1$ minimization problem is the classic least absolute deviations problem; Boyd~\cite{boyd2010distributed} has a nice survey of algorithms to solve this problem and describes  in detail a fast solver based on the technique of ADMM (Alternating Direction Method of Multipliers)\footnote{\url{ http://www.stanford.edu/~boyd/papers/admm/}}.  More references can be found therein.

According to~\cite{boyd2010distributed}, we can rewrite the right hand of Equation \eqref{eq:L1_distance} as the equivalent constrained problem by introducing a sparse outlier vector $s$:
\begin{eqnarray}\label{eq:L1_constrain}
	\min &&{\|  s \|_1} \\
	s.t. && U_{\Omega_t}w + s - v_{\Omega_t} = 0 \nonumber
\end{eqnarray}
The augmented Lagrangian of this constrained minimization problem is then

\begin{equation} \label{eq:L1_aug_Lagrangian}
	\mathcal{L} (s, w, y) = \| s \|_1 + y^T(U_{\Omega_t} w + s - v_{\Omega_t}) + \frac{\rho}{2}\| U_{\Omega_t} w + s - v_{\Omega_t} \|_2^2
\end{equation}
where $y$ is the dual vector. Our unknowns are $s$, $y$, $U$, and $w$. Note that since $U$ is constrained to a non-convex manifold ($U^TU=I$), this function is not convex (neither is Equation~\eqref{eq:grousecost}). However, note that if $U$ were estimated, we could solve for the triple $(s,w,y)$ using ADMM; also if $(s,w,y)$ were estimated, we could refine our estimate of $U$. This is the alternating approach we take with GRASTA. We describe the two parts in detail in Sections~\ref{sec:updateswy} and~\ref{sec:subspaceupdate}.

\subsection{Relation to Robust PCA and Robust Matrix Completion}
\label{sec:relation-rpca-rmc}

If the subspace $\mathcal{S}$ does not evolve over time, this problem reduces to subspace estimation, which can be related to Robust PCA. For a set of time samples $t=1, \dots, T$, we observe a sequence of incomplete corrupted data vectors $v_{\Omega_1}, \ldots, v_{\Omega_T}$. Let the matrix $V = \left[  v_1, \ldots, v_T  \right]$. Let $\mathcal{P}_{\Omega}( \cdot )$ denote operator which selects from each column the corresponding indices in $\Omega_1, \dots, \Omega_T$; thus $\mathcal{P}_\Omega(V)$ denotes our partial observation of the corrupted matrix $V$. Note that from our model in Equation~\eqref{eq:measmodel}, we can write $V$ as a sum of a sparse matrix $S$ and a low-rank matrix $L=UW$, where the orthonormal columns of $U \in \R^{n \times d}$ span $\Ss$ (which is stationary), and $W \in \R^{d \times T}$ holds the weight vectors $w_t$ as columns.

The global version of the $l^1$ cost function in Equation \eqref{eq:L1_distance} follows:

\begin{eqnarray} \label{eq:robust-mc}
	\bar{F}(\Ss) = \sum_{t=1}^{T} \min_w \| U_{\Omega_t}w - v_{\Omega_t} \|_1 
					&=& \min_{W\in \mathbb{R}^{d \times T}} \sum_{(i,j)\in \Omega}  | (UW - V)_{ij} | \\ \nonumber
					&=& \min_{W\in \mathbb{R}^{d \times T}} \| \mathcal{P}_{\Omega}( UW - V )\|_1 \;\;.
\end{eqnarray}
The right hand of Equation \eqref{eq:robust-mc} can be rewritten as the equivalent constrained problem:
\begin{eqnarray} \label{eq:rmc-constrain}
	\min &&\| \mathcal{P}_{\Omega}( S )\|_1 \\ \nonumber
	s.t.  && \mathcal{P}_{\Omega}  ( UW + S )  = \mathcal{P}_{\Omega} (V) \\ \nonumber
	       && U \in \mathcal{G}(d,n)
\end{eqnarray}
which is the same problem studied in~\cite{shen2011augmented}, and the authors propose an efficient ADMM solver for this problem. Unlike the set-up of~\cite{chandrasekaran2011rank,CandesRPCA09}, this problem is not convex; however it offers much more computationally efficient solutions. GRASTA differs from the algorithm of~\cite{shen2011augmented} in two major ways: it uses incremental gradient to minimize this cost function one column at a time for even greater efficiency, and it uses geodesics on the Grassmannian to compute the update of $U$.

% GLAD main body

\section{Grassmannian Robust Adaptive Subspace Tracking} 
\label{sec:alg_derivation}

As we have said, GRASTA alternates between estimating the triple $(s,w,y)$ and the subspace $U$. Here we discuss those two pieces of our algorithm. Section~\ref{sec:updateswy} describes the update of $(s,w,y)$ based on an estimate $\widehat{U}_t$ for the subspace variable. Section~\ref{sec:subspaceupdate} describes the update of our subspace variable to $\widehat{U}_{t+1}$ based on the estimate of $(s^*,w^*,y^*)$ resulting from the first step. Finally, Section~\ref{sec:adaptive_step} describes our algorithm for adaptively choosing the gradient step-size.

\subsection{Update of the sparse vector, weight vector, and dual vector}
\label{sec:updateswy}

Given the current estimated subspace $\widehat{U}_t$, the partial observation $v_{\Omega_t}$, and the observed entries' indices $\Omega_t$, the optimal  $(s^*, w^*, y^*) $ of Equation \eqref{eq:L1_constrain} can be found with the following minimization of the augmented Lagrangian.

\begin{equation} \label{eq:min_aug_Lagrangian}
 (s^*, w^*, y^*) = \arg\min_{s,w,y} \mathcal{L}(\widehat{U}_{\Omega_t}, s, w, y)
\end{equation}

Equation \eqref{eq:min_aug_Lagrangian} can be efficiently solved by ADMM~\cite{boyd2010distributed}. That is, $s$, $w$, and the dual vector $y$ are updated in an alternating fashion:
\begin{equation} \label{eq:L1_ADMM}
\left\{\begin{array}{l}
w^{k+1} = \arg\min_w\mathcal{L}( \widehat{U}_{\Omega_t}, s^k, w, y^k) \\
s^{k+1} = \arg\min_s\mathcal{L}( \widehat{U}_{\Omega_t}, s, w^{k+1}, y^k)  \\ 
 y^{k+1} =  y^k + \rho ( \widehat{U}_{\Omega_t} w^{k+1} + s^{k+1} - v_{\Omega_t})
\end{array} \right.
\end{equation}
%We write the ADMM detail below. 
Specifically, these quantities are computed as follows. In this paper we always assume that $U_{\Omega_t}^T U_{\Omega_t}$ is invertible, which is guaranteed if $|\Omega_t|$ is large enough~\cite{Balzano10a}. We have:

\begin{eqnarray}
w^{k+1} &=& \frac{1}{\rho}( \widehat{U}_{\Omega_t}^T \widehat{U}_{\Omega_t})^{-1} \widehat{U}_{\Omega_t}^T (\rho(v_{\Omega_t} - s^k) - y^k) \\
s^{k+1} &=& \mathsf{S}_{\frac{1}{1+\rho}} (v_{\Omega_t} - \widehat{U}_{\Omega_t} w^{k+1} - y^k )  \\ 
y^{k+1} &=&  y^k + \rho ( \widehat{U}_{\Omega_t} w^{k+1} + s^{k+1} - v_{\Omega_t})
\end{eqnarray}
where $ \mathsf{S}_{\frac{1}{1+\rho}}$ is the elementwise soft thresholding operator~\cite{boyd2004convex}. We discuss this ADMM solver in detail as Algorithm \ref{alg:SRP} in Section \ref{sec:algos}.

\subsection{Subspace Update}
\label{sec:subspaceupdate}

As we mentioned in Section \ref{sec:background}, the set of all subspaces of  $\mathbb{R}^n $ of fixed dimension $d$ is called \textit{Grassmannian}, which is a compact Riemannian manifold, and is denoted by $\mathcal{G}(d,n)$. Edelman, Arias and Smith (1998) have a comprehensive survey~\cite{Edelman98} that covers how both the Grassmannian geodesics and the gradient of a function defined on the Grassmannian manifold can be explicitly computed. 

\begin{figure}[h!]
	\begin{center}
		\includegraphics[width=0.4\textwidth]{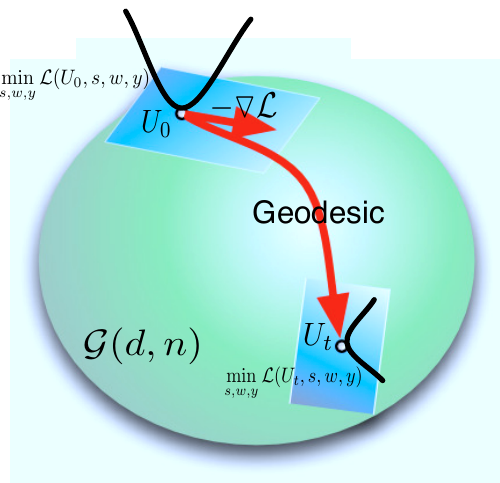}
		\caption{Illustration of the gradient descent along geodesic on the Grassmannian manifold}
		\label{fig:glad_demo}
	\end{center}	
\end{figure}

GRASTA achieves online robust subspace tracking by performing incremental gradient descent on the Grassmannian step by step. That is, we first compute a gradient of the loss function, and then follow this gradient along a short geodesic curve on the Grassmannian. Figure \ref{fig:glad_demo} illustrates the basic idea of gradient descent along a geodesic. 

\subsubsection{Augmented Lagrangian as the Loss Function}
It seems that it would be natural to use Equation \eqref{eq:L1_distance} as the robust loss function. However, there is a critical limitation of this approach: when regarding $U$ as the variable, this loss function is not differentiable everywhere.

Here we propose to use the augmented Lagrangian as the subspace loss function once we have estimated $(s^*, w^*, y^*)$ from the previous $\widehat{U}_{\Omega_t}$ and $v_{\Omega_t}$ by Equation \eqref{eq:L1_ADMM}. The new loss function is stated as Equation \eqref{eq:loss_aug_Lagrangian}:

\begin{equation}
	\label{eq:loss_aug_Lagrangian}
\mathcal{L}(U) = \| s^* \|_1 + {y^*}^T( U_{\Omega_t} w^*+s^*-v_{\Omega_t}) + \frac{\rho}{2} \| U_{\Omega_t}  w^*+s^*-v_{\Omega_t} \|_2^2 \\
\end{equation}
This new subspace loss function is differentiable. Furthermore, when the data vector is not corrupted by outliers, Equation \eqref{eq:loss_aug_Lagrangian} reduces to the $l^2$-norm loss function of GROUSE~\cite{grouse}.

\subsubsection{Grassmannian Geodesic Gradient Step}
In order to take a gradient step along the geodesic of the Grassmannian, according to~\cite{Edelman98}, we first need to derive the gradient formula of the real-valued loss function Equation \eqref{eq:loss_aug_Lagrangian} $\mathcal{L}: \mathcal{G}(d,n) \rightarrow \mathbb{R}$.

From Equation (2.70) in~\cite{Edelman98}, the gradient $\triangledown{\mathcal{L}} $ can be determined from the derivative of $\mathcal{L}$ with respect to the components of $U$.  
Let $\chi_{\Omega_t}$ is defined to be the $|\Omega_t|$ columns of an $n\times n$ identity matrix corresponding to those indices in $\Omega_t$; that is, this matrix zero-pads a vector in $\R^{|\Omega_t|}$ to be length $n$ with zeros on the complement of $\Omega_t$. 
The derivative of the augmented Lagrangian loss function $\mathcal{L}$ with respect to the components of $U$ is as follows:

\begin{equation} \label{eq:derivative}
	\frac{d \mathcal{L}}{d U} = \left[ \chi_{\Omega_t} \left( y^* + \rho (U_{\Omega_t} w^* + s^* - v_{\Omega_t} ) \right) \right] {w^*}^T
\end{equation}
Then the gradient $\triangledown{\mathcal{L}}$ is $\triangledown{\mathcal{L}}  = (I - UU^T) \frac{d \mathcal{L}}{d U}$~\cite{Edelman98}.
%
%\begin{eqnarray} \label{eq:gradient}
%	\triangledown{\mathcal{L}}  &=& (I - UU^T) \frac{d \mathcal{L}}{d U}  \\   \nonumber
%												&=& (I - UU^T)  \left[ y^* + \rho (\bigtriangleup_{\Omega_t} U w^* + s^* - v_{\Omega_t} )  \right] {w^*}^T 
%\end{eqnarray}
Here we introduce three variables $\Gamma$, $\Gamma_1$, and $\Gamma_2$ to simplify the gradient expression:

\begin{eqnarray}
	\Gamma_1 &=& y^* + \rho (U_{\Omega_t} w^* + s^* - v_{\Omega_t} ) \\
	\Gamma_2 &=& U_{\Omega_t}^T \Gamma_1 \\
	\Gamma   &=& \chi_{\Omega_t}\Gamma_1 - U \Gamma_2 
\end{eqnarray}
Thus the gradient $\triangledown{\mathcal{L}} $ can be further simplified to:

\begin{equation} \label{eq:gradient_simp}
	\triangledown{\mathcal{L}}  = \Gamma {w^*}^T
\end{equation}

From Equation \eqref{eq:gradient_simp}, it is easy to verify that $\triangledown{\mathcal{L}} $ is rank one since $\Gamma$ is a $n \times 1$ vector and $w^*$ is the optimal $d \times 1$ weight vector. Then it is trivial to compute the singular value decomposition of  $\triangledown{\mathcal{L}}$, which will be used for the following gradient descent step along the geodesic according to Equation (2.65) in~\cite{Edelman98}. The sole non-zero singular value is $\sigma = \|\Gamma \| \| w^* \|$, and the corresponding left and right singular vectors are $\frac{\Gamma}{\| \Gamma \|}$ and $\frac{w^*}{\|w^* \|}$ respectively. Then we can write the SVD of the gradient explicitly by adding the orthonormal set $x_2, \ldots, x_d$ orthogonal to $\Gamma$ as left singular vectors and the orthonormal set $y_2, \ldots, y_d$ orthogonal to $w^*$ as right singular vectors as follows:

\begin{equation} \label{eq:svd_grad}
	 \triangledown{\mathcal{L}}  = \left[ \frac{\Gamma}{\| \Gamma \|} ~~ x_2~~\ldots ~~ x_d \right] \times diag(\sigma, 0, \ldots, 0) \times \left[ \frac{w^*}{\|w^* \|} ~~ y_2 ~~\ldots ~~ y_d \right]^T
\end{equation}

Finally, following Equation (2.65) in~\cite{Edelman98}, a gradient step of length $\eta$  in the direction $- \triangledown{\mathcal{L}}$ is given by 

\begin{equation} \label{eq:gradient_step}
	U(\eta) = U + \left( (cos(\eta \sigma)-1) \frac{Uw_t^*}{\|w_t^*\|}  - sin(\eta \sigma)\frac{\Gamma}{\|\Gamma\|} \right) \frac{ {w_t^*}^T}{\|w_t^*\|} \;.
\end{equation}

\subsection{Remarks}
Here we point out that at each subspace update step, our approach does not remove outliers explicitly. In fact, we use the gradient of the augmented Lagrangian $\mathcal{L}(U)$ Equation \eqref{eq:loss_aug_Lagrangian} which exploits the dual vector $y^*$ to leverage the outlier effect. That is the key to success. Even when the ADMM solver \ref{eq:L1_ADMM} can not identify the outliers due to our current estimated subspace being far away from the true subspace, with the help of the dual vector $y^*$ the gradient of the augmented Lagrangian gives us the right direction at each step which leads us to the right subspace. 

We also must point out that since we estimate $(s^*, w^*, y^*) $ at each step using the ADMM solver, we can not recover the exact subspace with sufficient accuracy if we do not allocate enough iterations for the ADMM solver~\cite{boyd2010distributed}. Fortunately, as it also emphasized in~\cite{boyd2010distributed}, only a few tens of iterations per subspace update step are sufficient to achieve a modest accuracy, which is often acceptable for practical use.  Extensive experiments in Section \ref{sec:expts} show that our algorithm is fast and always produces acceptable results, even when the vectors are noisy and heavily corrupted by outliers.

\subsection{Adaptive Step-size} 
\label{sec:adaptive_step}

The question of how large a gradient step to take along the geodesic is an important issue, and it depends on a fundamental tradeoff between tracking rate and steady-state error.  Rather than the constant step-size rule proposed for subspace tracking in GROUSE, here we propose to use the adaptive step-size rule to achieve both precise convergence for a stationary subspace and fast adaptation to a changing subspace.

We use the following formula to update the step-size $\eta_t$:

\begin{equation} \label{eq:step_size}
	\eta_t = \frac{C}{1+ \mu_t}
\end{equation}
where $C$ is the predefined constant step-size scale. 
If we use $\mu_t = t$ to update $\eta_t$, it is obvious that the step-size satisfies the following properties:
		$$ \lim_{t\rightarrow \infty } \eta_t = 0 \qquad  \textrm {and} \qquad \sum_{t=1}^{\infty} \eta_t = \infty$$
This is the classic diminishing step-size rule in stochastic gradient descent literature, and has been proven to guarantee convergence to a stationary point~\cite{robbins1951stochastic}~\cite{kushner2003stochastic}.

However, our goal is not only to identify the stationary subspace precisely. We have the more ambitious goal of keeping track of the subspace when the subspace is slowly changing. Obviously, with a changing subspace, if we use a diminishing step-size rule, when $\eta_t$ is shrinking to $0$ our steps will be too small to track the dynamic subspace.  To continually adapt to the changing subspace, GROUSE~\cite{grouse} proposes a constant step-size which needs careful selection to balance the tradeoff between tracking rate and steady-state error. 

Here we propose to use an adaptive step-size rule to produce a proper step-size $\eta_t$ that empirically achieves both precise convergence for a stationary subspace and fast adaptation to a changing subspace. The basic idea is inspired by Plakhov~\cite{plakhov2004stochastic} and Klein~\cite{klein2009adaptive}: if two consecutive gradients $\triangledown \mathcal{L}_{t-1}$ and $\triangledown \mathcal{L}_{t}$ are in the same direction, i.e. $\langle \triangledown \mathcal{L}_{t-1},\triangledown \mathcal{L}_{t} \rangle  > 0$, it intuitively means that the current estimated $\widehat{U}_t$ is relatively far away from the true subspace $\mathcal{S}_t$. If this is the case, heuristically we should take a slightly larger step along $-\triangledown \mathcal{L}_{t}$ than the previous step-size $\eta_{t-1}$. Otherwise, if $\triangledown \mathcal{L}_{t-1}$ and $\triangledown \mathcal{L}_{t}$ are not in the same direction, i.e. $\langle \triangledown \mathcal{L}_{t-1},\triangledown \mathcal{L}_{t} \rangle  < 0$, again intuitively this means that the current estimated $\widehat{U}_t$ is relatively close the true subspace $\mathcal{S}_t$, and again heuristically we should take a slightly smaller step along $-\triangledown \mathcal{L}_{t}$ than the previous step-size $\eta_{t-1}$. Besides the sign of the two consecutive gradients giving us intuition for the step-size adaptation, the inner product $\langle \triangledown \mathcal{L}_{t-1},\triangledown \mathcal{L}_{t} \rangle $ also gives us the proper adapted magnitude for our step-size ~\cite{plakhov2004stochastic}~\cite{klein2009adaptive}.

We still use Equation \eqref{eq:step_size} to produce $\eta_t$ at each time, but update $\mu_t$ according to the inner product of two consecutive gradients $\langle \triangledown \mathcal{L}_{t-1},\triangledown \mathcal{L}_{t} \rangle$ as follows:
\begin{equation} \label{eq:step_size_variable}
	\mu_t = \max{  \{\mu_{t-1} + sigmoid(-\langle \triangledown \mathcal{L}_{t-1},\triangledown \mathcal{L}_{t} \rangle), 0 \}  } 
\end{equation}
where the $sigmoid$ function is defined as:
		$$ sigmoid(x) = f_{MIN}+\frac{f_{MAX} - f_{MIN}}{1-(f_{MAX}/f_{MIN})e^{-x/\omega}} $$
with $sigmoid(0)=0$, $f_{MAX}>0$, $ f_{MIN}<0$, and $\omega > 0$. $f_{MAX}$ and $ f_{MIN}$ are chosen to control how much the step-size grows or shrinks; and $\omega$ controls the shape of the $sigmoid$ function. In this paper we always set $f_{MAX}=1$, $ f_{MIN}=-1$, and  $\omega = 0.1$.

When the estimated subspace $\widehat{U}_t$ is very close to the true  subspace $\Ss_t$, the adaptive step-size $\eta_t \rightarrow 0$, or equivalently $\mu_t \rightarrow +\infty$ from Equation \eqref{eq:step_size}. Now we consider the following scenario: suppose we have identified the subspace precisely-- and therefore $\mu_t > N$ for some large number $N$
then suddenly the subspace changes dramatically. How quickly will this step-size rule adapt to the new subspace? 
In practical applications, taking too much time to adapt to the new subspace is undesirable. Specifically, only shrinking $\mu_t$ at most $|f_{MIN}|$ is too conservative in this scenario. It is easy to verify that, since at each update step $\mu_t$ shrinks at most $|f_{MIN}|$, the increase of $\eta_t$ is limited and therefore this approach wouldn't take very large steps even though the subspace has changed. When the subspace changes drastically, we should shrink $\mu_t$ more to accelerate the adaptation process. 

For GRASTA, we take this approach and call it a "Multi-Level" adaptive step-size rule. Though we do not provide the convergence proof here, empirically this multi-level adaptive approach demonstrates much faster convergence performance than the single-level strategy discussed above. We leave further detailed comparison to future investigation.

Our multi-level adaptation is as follows. We only let $\mu_t$ change in $\left( \mu_{MIN}, \mu_{MAX} \right)$, where $\mu_{MIN}$ and $\mu_{MAX}$ are prescribed constants. For the experiments in this paper we always set $\mu_{MIN}=1$ and $\mu_{MAX}=15$. Then in this case Equation \eqref{eq:step_size_variable} is adapted to Equation \eqref{eq:step_size_variable_range}:

\begin{equation} \label{eq:step_size_variable_range}
	\mu_t = \max{  \{\mu_{t-1} + sigmoid(-\langle \triangledown \mathcal{L}_{t-1},\triangledown \mathcal{L}_{t} \rangle), \mu_{MIN} \}  } 
\end{equation}
We introduce a level variable $l_t$ that will get smaller when our subspace estimate is far from the data. Then the step-size $\eta_t$ is as follows:

\begin{equation} \label{eq:multi-level-adaptive}
	\eta_t = \frac{C2^{-l_t}}{1+ \mu_t}
\end{equation}

Once $\mu_t$ calculated by Equation \eqref{eq:step_size_variable_range} is larger than $\mu_{MAX}$, we increase the level variable $l_t$ by $1$ and set $\mu_t=\mu_0$, where $\mu_0 \in \left( \mu_{MIN}, \mu_{MAX} \right)$ and $\mu_0$ is selected close to $\mu_{MIN}$ (in our experiments we let $\mu_0 = 3$). If $\mu_t \leq \mu_{MIN}$, we decrease $l_t$ by $1$ and also set $\mu_t=\mu_0$. Therefore, when our subspace estimate is off, we are increasing $\eta_t$ exponentially instead of linearly. On one hand this new multi-level adaptive rule follows the basic adaptive step-size rule discussed above; %but it also introduces an associated level information;
on the other hand exploiting this multi-level property, this new approach adapts more quickly to a changing subspace.  
Once we have identified the subspace changing and $\mu_t \leq \mu_{MIN}$, if the subspace really changes  dramatically, $l_t$ will keep decreasing until $\mu_t$ is again within the range $\left(\mu_{MIN}, \mu_{MAX} \right)$. 

Combining these ideas together, we state our novel adaptive step-size rule as Algorithm \ref{alg:Adaptive}.

\subsection{Algorithms}
\label{sec:algos}
The discussion of Sections~\ref{sec:updateswy} to~\ref{sec:adaptive_step} can be summarized into our algorithm as follows. For each time step $t$, when we observe an incomplete and corrupted data vector $v_{\Omega_t}$, our algorithm will first estimate the optimal value  $ (s^*, w^*, y^*) $ from our current estimated subspace $U_t$ via the $l^1$ minimization ADMM solver \ref{eq:L1_ADMM}; then compute the gradient of the augmented Lagrangian loss function $\mathcal{L}$ by Equation \eqref{eq:gradient_simp};  then estimate a proper step-size $\eta_t$ from the two consecutive
gradients $\triangledown \mathcal{L}_{t-1} $ and $\triangledown
\mathcal{L}_t $ by Equation \eqref{eq:step_size_variable} and
\ref{eq:multi-level-adaptive} ; and finally do the rank one subspace update via Equation \eqref{eq:gradient_step}. 

We state our main algorithm GRASTA (Grassmannian Robust Adaptive Subspace Tracking Algorithm) in Algorithm \ref{alg:GLAD}. GRASTA consists of two important sub-procedures: the ADMM solver of the least absolute derivations problem, and the computation of the adaptive step-size. We state the two sub-procedures as Algorithm \ref{alg:SRP} and Algorithm \ref{alg:Adaptive} separately.

\begin{algorithm}
\caption{Grassmannian Robust Adaptive Subspace Tracking}\label{alg:GLAD}
\textbf{Require}: An $n\times d$ orthogonal matrix $U_0$. A sequence of corrupted vectors $v_t$, each vector observed in entries $\Omega_t \subset \{1, \dots, n\}$. A structure OPTS1 that holds parameters for ADMM. A structure OPTS2 that holds parameters for the adaptive step size computation.

\textbf{Return}: The estimated subspace $U_t$ at  time  $t$.

\begin{algorithmic}[1]
%\STATE Initialize $U_1$: $U_1 = U_0$
\FOR {$t=0,\ldots,T$}
\STATE Extract $U_{\Omega_t}$ from $U_t$: $U_{\Omega_t} = \chi_{\Omega_t}^T U_t$ 
\STATE Estimate the sparse residual $s_t^*$, weight vector $w_t^*$, and dual vector $y_t^*$ from the observed entries $\Omega_t$ via Algorithm \ref{alg:SRP} using OPTS1: \\
$\qquad\qquad\qquad$$(s_t^*, w_t^*, y_t^*) =  \arg\min_{w,s,y}\mathcal{L}(U_{\Omega_t}, w, s, y) $\\ %$
\STATE Compute the gradient of the augmented Lagrangian $\mathcal{L}$, $\triangledown{\mathcal{L}} $ as follows:\\
 $\qquad$ $\Gamma_1 = y_t^* + \rho (U_{\Omega_t} w_t^* + s_t^* - v_{\Omega_t})$,  
 $\quad$	 $\Gamma_2 = U_{\Omega_t}^T \Gamma_1$, 
 $\quad$  $\Gamma = \chi_{\Omega_t} \Gamma_1 - U \Gamma_2$ \\
 $\qquad \qquad \qquad$ $\triangledown{\mathcal{L}} = \Gamma {w_t^*}^T$
 \STATE Compute step-size $\eta_t$ via the adaptive step-size update rule according to Algorithm \ref{alg:Adaptive} using OPTS2.
 \STATE Update subspace: $U_{t+1} =  U_t + ((cos(\eta_t \sigma)-1)U_t\frac{w_t^*}{\|w_t^*\|}  - sin(\eta_t \sigma)\frac{\Gamma}{\|\Gamma\|}) \frac{ {w_t^*}^T}{\|w_t^*\|}$\\
 $\qquad \qquad \qquad$  where $\sigma = \|\Gamma\| \|w_t^*\|$  %$
 
\ENDFOR
\end{algorithmic}

\end{algorithm}

Unlike GROUSE, which has a closed form solution for computing the gradient, GRASTA estimates $(s_t^*, w_t^*, y_t^*) $ by the ADMM iterated Algorithm \ref{alg:SRP}. Certainly we would have a potential performance bottleneck if Algorithm \ref{alg:SRP} takes too much time at each subspace update step. However, we see empirically that only a few tens of iterations in Algorithm \ref{alg:SRP} at each step allows GRASTA to track the subspace to an acceptable accuracy. In our video experiments with Algorithm \ref{alg:SRP}, we always set the maximum iteration $K$ around $20$ to balance the trade-off between the subspace tracking accuracy and computational performance. We make a slight modification to the original ADMM sovler presented in~\cite{boyd2010distributed}: in addition to returning $w^*$ we also return the sparse vector $s^*$ and the dual vector $y^*$ for the further computation of the gradient $\triangledown{\mathcal{L}} $. It is easy to verify that  in the worst case the ADMM solver needs at most $O(|\Omega|d^3 + Kd|\Omega| )$ flops.
 
\begin{algorithm}
\caption{ADMM Solver for Least Absolute Deviations~\cite{boyd2010distributed}}\label{alg:SRP}
\textbf{Require}:  An $|\Omega_t|\times d$ orthogonal matrix $U_{\Omega_t}$, a corrupted observed vector $v_{\Omega_t} \in \R^{|\Omega_t|}$ ,  and a structure OPTS which holds four parameters for ADMM: ADMM step-size constant $\rho$, the absolute tolerance $\epsilon^{abs}$, the relative tolerance $\epsilon^{rel}$, and ADMM maximum iteration $K$.

\textbf{Return}: sparse residual $s^*\in \R^{|\Omega_t|}$; weight vector $w^*\in \R^d$; dual vector $y^*\in \R^{|\Omega_t|}$.

\begin{algorithmic}[1]
\STATE Initialize $s$,$w$,$y$: $s^1=s^0$, $w^1=w^0$, $y^1=y^0$\\(either to zero or to the final value from the last subspace update of the same data vector for a warm start.)
\STATE Cache $P = (U_{\Omega_t}^T U_{\Omega_t})^{-1}U_{\Omega_t}^T$
\FOR{$k = 1 \to K$ } 
\STATE Update weight vector: $w^{k+1} = \frac{1}{\rho} P (\rho(v_{\Omega_t} - s^k) - y^k)$ 
\STATE Update sparse residual: $s^{k+1} = \textsf{S}_{\frac{1}{\rho+1}} (v_{\Omega_t} - U_{\Omega_t} w^{k+1} - y^k) $
\STATE Update dual vector: $y^{k+1} = y^k + \rho (U_{\Omega_t} w^{k+1} + s^{k+1} - v_{\Omega_t})$
\STATE Calculate primal and dual residuals: $r^{pri} = \| U_{\Omega_t} w^{k+1} + s^{k+1} - v_{\Omega_t}\|$, 
$ r^{dual}  = \|\rho  U_{\Omega_t}^T  (s^{k+1} - s^k)\|$
\STATE Update stopping criteria: $\epsilon^{pri}=\sqrt{|\Omega_t|}\epsilon^{abs} + \epsilon^{rel}\max{\{\|U_{\Omega_t} w^{k+1}\|, \|s^{k+1}\|, \|v_{\Omega_t}\| \}}$, \\
				$\qquad\qquad\qquad\qquad\qquad\quad$			$\epsilon^{dual}=\sqrt{d}\epsilon^{abs}+ \epsilon^{rel}\|\rho U_{\Omega_t}^Ty^{k+1}\|$
\IF {$r^{pri} \leq \epsilon^{pri}$ \AND $r^{dual} \leq \epsilon^{dual}$}
\STATE Converge and break the loop.
\ENDIF
\ENDFOR
\STATE $s^*=s^{K+1}$, $w^* = w^{K+1}$, $y^* = y^{K+1}$
\end{algorithmic}

\end{algorithm}

%%%%%$

In order to produce the proper step-size $\eta_t$ from Algorithm \ref{alg:Adaptive}, we need to maintain the gradient $\triangledown \mathcal{L}_{t-1}$ from the previous time step throughout the subspace tracking process. Keeping $\triangledown \mathcal{L}_{t-1}$ only requires additional $O(n+d)$ memory usage. The main computation of of Algorithm \ref{alg:Adaptive} is the inner product $ \langle \triangledown \mathcal{L}_{t-1},\triangledown \mathcal{L}_{t} \rangle$,  which is the trace of the product $\triangledown \mathcal{L}_{t-1}$ and $\triangledown \mathcal{L}_{t}$, two $n \times d$ matrices, and will cost $O(nd^2)$ flops.

\begin{algorithm}
\caption{Multi-Level Adaptive Step-size Update}\label{alg:Adaptive}
\textbf{Require}:  Previous gradient $\triangledown \mathcal{L}_{t-1}$ at time $t-1$,  current gradient  $\triangledown \mathcal{L}_{t}$ at time $t$.  Previous step-size variable $\mu _{t-1}$. Previous level variable $l_{t-1}$. Constant step-size scale $C$. Adaptive step-size parameters $f_{MAX}, f_{MIN}, \mu_{MAX}, \mu_{MIN}$.

\textbf{Return}: Current step-size $\eta_t$,  step-size variable $\mu_t$, and level variable $l_t$.

\begin{algorithmic}[1]
\STATE Update the step-size variable: $\mu_t = \max{  \{\mu_{t-1} + sigmoid(-\langle \triangledown \mathcal{L}_{t-1},\triangledown \mathcal{L}_{t} \rangle), \mu_{MIN} \}  }$\\
where $sigmoid$ function is defined as: \\
$\qquad\qquad\qquad\qquad$ $sigmoid(x) = f_{MIN}+\frac{f_{MAX} - f_{MIN}}{1-(f_{MAX}/f_{MIN})e^{-x/\omega}}$, with $sigmoid(0)=0$.
\IF {$\mu_t \geq \mu_{MAX}$}
\STATE Increase to a higher level: $l_t=l_{t-1}+1$ and $\mu_t = \mu_0$
\ELSIF {$\mu_t \leq \mu_{MIN}$}
\STATE Decrease to a lower level: $l_t = l_{t-1} -1$ and $\mu_t = \mu_0$
\ELSE
\STATE Keep at the current level: $l_t = l_{t-1}$
\ENDIF
\STATE Update the step-size: $\eta_t = C2^{-l_t}/(1+ \mu_t)$
\end{algorithmic}
\end{algorithm}

\subsection{Computational Cost and Memory Usage}

Each subspace update step in GRASTA needs only simple linear algebraic computations. The total computational cost of each step of Algorithm~\ref{alg:GLAD} is  $O(|\Omega|d^3 + Kd|\Omega| + nd^2)$, where again $|\Omega|$ is the number of samples per vector used, $d$ is the dimension of the subspace, $n$ is the ambient dimension, and $K$ is the number of ADMM iterations. 

Specifically, estimating $(s_t^*, w_t^*, y_t^*) $ from Algorithm \ref{alg:SRP} costs at most $O(|\Omega|d^3 + Kd|\Omega| )$ flops; computing the gradient $\triangledown \mathcal{L}$ needs simple matrix-vector multiplication which costs $O(|\Omega|d + nd)$ flops; producing the adaptive step-size costs $O(nd^2)$ flops; and the final update step also costs $O(nd^2)$ flops. 

Throughout the tracking process, GRASTA only needs $O(nd)$ memory elements to maintain the estimated low-rank orthonormal basis $\widehat{U}_t$, $O(n)$ elements for $s^*$ and $y^*$, $O(d)$ elements for $w^*$, and $O(n+d)$ for the previous step gradient $\triangledown \mathcal{L}_{t-1}$ in memory. 

This analysis decidedly shows that GRASTA is both computation and memory efficient.

% Numerical Experiments

\section{Numerical Experiments}
\label{sec:expts}

In the following experiments, we explore GRASTA's performance in various scenarios: subspace tracking, robust matrix completion, and the video surveillance application. We use relative error to quantify the performance of GRASTA. If the recovered data is a vector, the relative error is defined as follows:
\begin{equation}
	RelErr =\frac{ \|\widehat{v} - v \|_2 }{\| v \|_2} 
\end{equation}
If the recovered data is a matrix, the relative error is defined as follows:
\begin{equation}
	RelErr = \frac{ \|\widehat{M} - M \|_F }{\| M \|_F} 
\end{equation}
We also use "Noise Relative Power" to quantify the additional Gaussian white noise perturbation, which is defined as follows:
\begin{equation}
	N_{Rel} = \frac{\| \zeta \|_2}{\| v\|_2}
\end{equation}
Here $v$ is the true data vector and $\zeta$ is the additional Gaussian noise as in Equation~\eqref{eq:measmodel}.

In all the following experiments, we use Matlab R2010b on a Macbook Pro laptop with 2.3GHz Intel Core i5 CPU and 4 GB RAM. To improve the performance, we implement Algorithm \ref{alg:SRP} in C++ and make it as a MEX-file to be integrated into GRASTA Matlab scripts.
\subsection{Comparison with GROUSE}
\label{sec:GROUSEcompare}

Our first goal is to compare GRASTA with the non-robust algorithm GROUSE to show the need for a robust subspace estimation and tracking algorithm.

\subsubsection{Subspace Tracking with Sparse Outliers}

In many of the following experiments,  we use this generative model to generate a series of data vectors:
\begin{equation} \label{eq:generative_model}
	v_t = U_{true}w_t + s_t + \zeta_t \;\;.
\end{equation}
$U_{true}$ is an $n \times d$ matrix whose $d$ columns are realizations of an i.i.d. $\mathcal{N}(0,I_n)$ random variable that are then orthornomalized. The weight vector $w_t$ is a $d \times 1$ vector whose entries  are realizations  of i.i.d. $\mathcal{N}(0,1)$ random variables, that is Gaussian distributed with mean zero and variance 1. The sparse vector $s_t$ is an $n \times 1$ vector whose nonzero entries are Gaussian noise with the maximum of the data vector $U_{true}w$ as the variance; the locations of the nonzero entries are chosen uniformly at random without replacement. The noise $\zeta_t$ is an $n \times 1$ vector whose  entries are i.i.d $\mathcal{N}(0, \omega^2)$. This parameter $\omega^2$ governs the SNR with respect to the low-rank part of our data. For the entire comparison against GROUSE, we used a maximum of $K=60$ iterations of the ADMM algorithm per subspace iteration.  

\begin{figure}[!h]
	\begin{center}
		\includegraphics[width=0.7\textwidth]{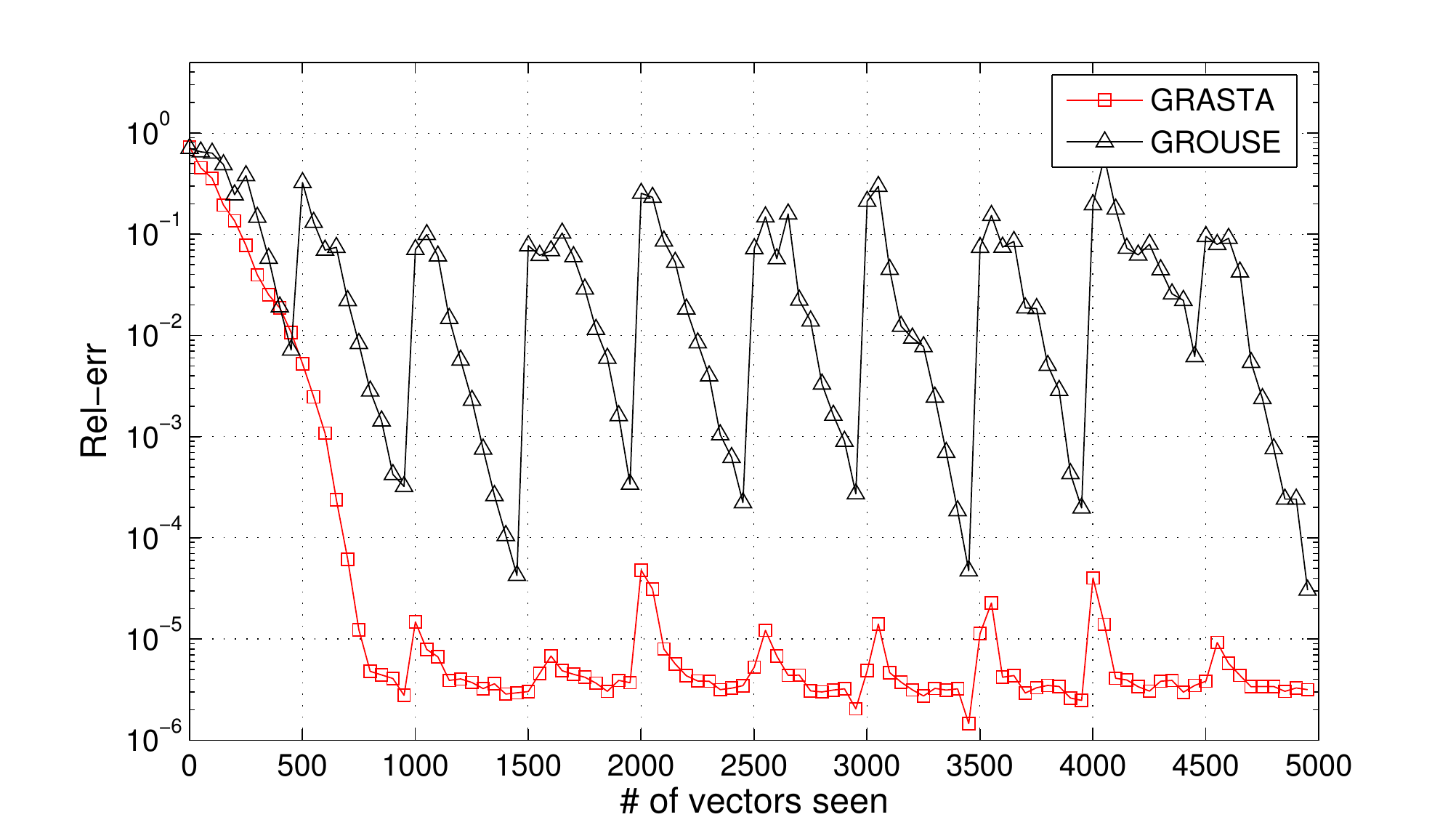} 
		\caption{Subspace tracking comparison between GROUSE and GRASTA from partial information. At time 500, 1000, $\ldots$, and 4500, 10\% observed entries are corrupted by outliers, and all entries are perturbed by small Gaussian noise with the variance of $\omega^2=10^{-6}$. }
		\label{fig:fig_cmp_grouse2}
	\end{center}	
\end{figure}

Figure \ref{fig:fig_cmp_grouse2} illustrates the failure of GROUSE, and success of GRASTA, when these sparse outliers are added only at periodic time intervals. We can see that GROUSE is significantly thrown off, despite the outliers occurring in an isolated vector. This illustrates clearly our motivation for adding robustness to the subspace tracking algorithm.

\subsubsection{Robust Matrix Completion}

We aim to complete $500 \times 500$ dimensional matrices of rank $5$. The matrices are corrupted by different fractions of outliers, depending on the experimental setting, and sampled uniformly without replacement with density $0.30$. We generate the low-rank matrix by first generating two $500 \times 5$ factors $Y_L$ and $Y_R$ with i.i.d. Gaussian entries and then adding normally distributed noise with variance $\omega^2$. 
The location of sparse outliers is distributed uniformly, and the outlier values are normally distributed with variance equal to the maximum of the matrix. 

For each setting of the fraction of outliers, we randomly generate $5$ matrices, each of which is solved via GROUSE and GRASTA separately. Both GROUSE and GRASTA cycle through the matrix columns $10$ times. Table \ref{tbl:cmp_grouse} shows the averaged results of a comparison between GROUSE and GRASTA. As expected, GRASTA vastly outperforms GROUSE across the board even with the smallest number of outliers.%on robust matrix completion. 

\begin{table}[!h]
\begin{center}
	\begin{tabular}{ c | c | c | c | c | c | c }
	\hline
						&\multicolumn{5}{c}{Fraction of Outliers}	\\ \cline{2-7}
						&  0				&	0.01		&	0.05		&	0.10		&	0.15		&	0.20		\\		\hline
		GROUSE	& 3.17E-6	&3.95E-1	&5.04E-1	&8.27E-1	&8.79E-1	&9.35E-1 \\		%\hline
		GRASTA	& 7.25E-6	&8.92E-5	&1.13E-4	&2.14E-4	&2.91E-4	&4.41E-4	\\		\hline
	\end{tabular}
\end{center}
	\caption{Robust matrix completion comparison between GRASTA and GROUSE. We only observe 30\%  of the low-rank matrix which is corrupted by sparse outliers. We show the averaged results of $5$ trials with different fractions of outliers. Here the matrix is $500 \times 500$, the rank is $5$, and all entries are perturbed by small Gaussian noise with the variance of $\omega^2 = 10^{-6}$. } 
		\label{tbl:cmp_grouse}
\end{table}

\subsection{Stationary Subspace Identification}

Now we wish to examine GRASTA's performance on the {\em stationary} 
subspace identification problem under various conditions. In most experiments (and unless otherwise noted) the ambient dimension is $n=500$ and the inherent subspace dimension is $d=5$. We again generate the vectors using Equation~\eqref{eq:generative_model} above and the descriptive text that follows Equation~\eqref{eq:generative_model}.
We vary the fraction of entries that are corrupted, and we vary the fraction of entries that are observed.

\begin{figure}[!h]
	\begin{center}
		\includegraphics[width=0.7\textwidth]{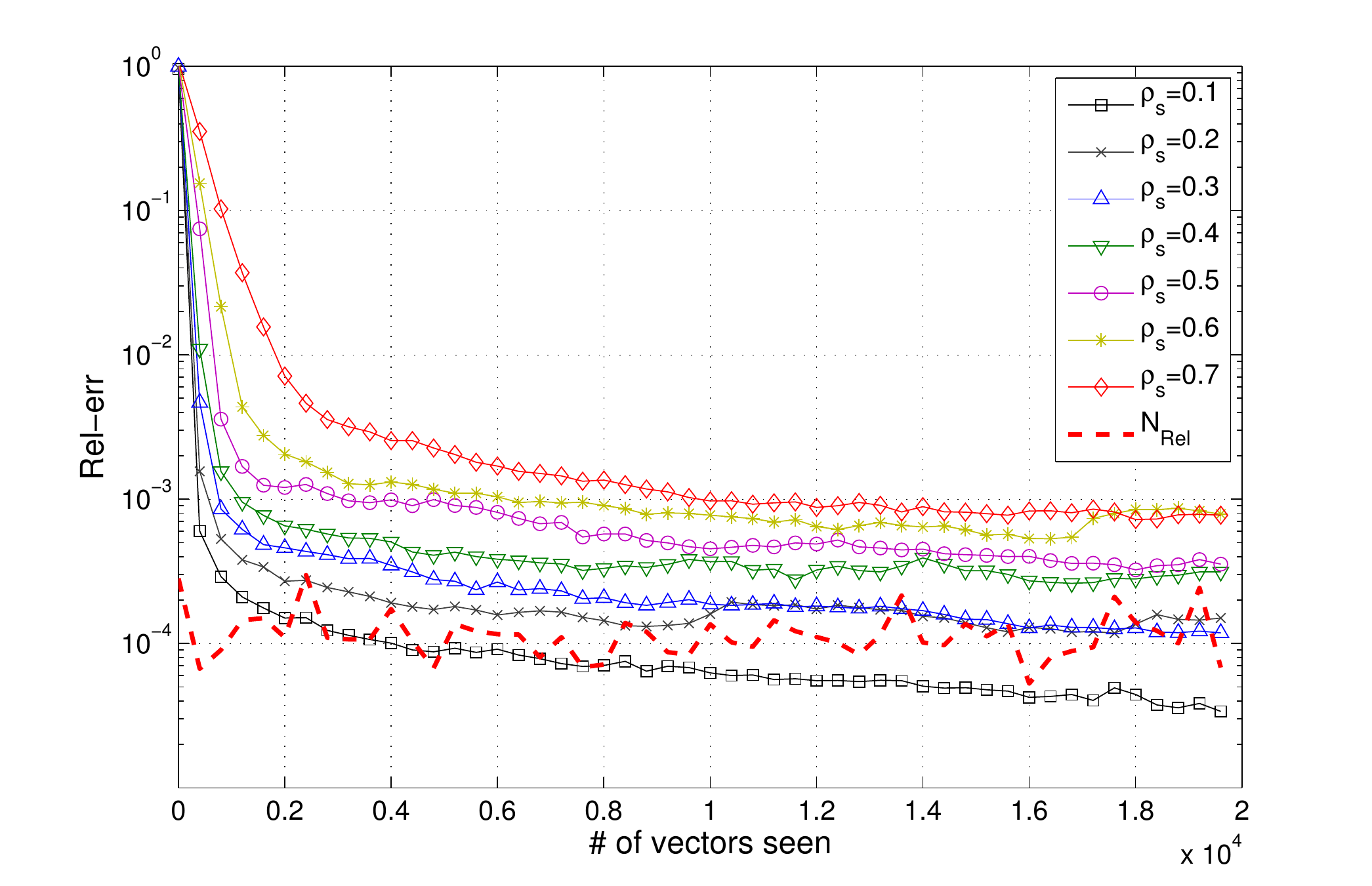}  
		\caption{The performance of stationary subspace identification  using full information within different fractions of outliers. We show the results from sparse outliers 10\% to dense outliers 70\%. The ambient dimension is $n=500$, and the subspace dimension is $d=5$. All observed entries are also perturbed by small Gaussian noise with the variance of $\omega^2 = 10^{-5}$.}
		\label{fig:full_var_spr}
	\end{center}	
\end{figure}

We start with Figure~\ref{fig:full_var_spr}, which shows subspace estimation performance under a varying fraction of added outliers. We can see in this problem instance that with 10\% corrupted entries, the relative error reaches the relative noise floor after a number of iterations that is a small multiple of the ambient dimension. For more corruption, more vectors (gradient iterations) are needed, but even with 50\% outliers and more, the relative error trends toward the relative noise power.

\begin{figure}[!h]
	\begin{center}
		\includegraphics[width=0.7\textwidth]{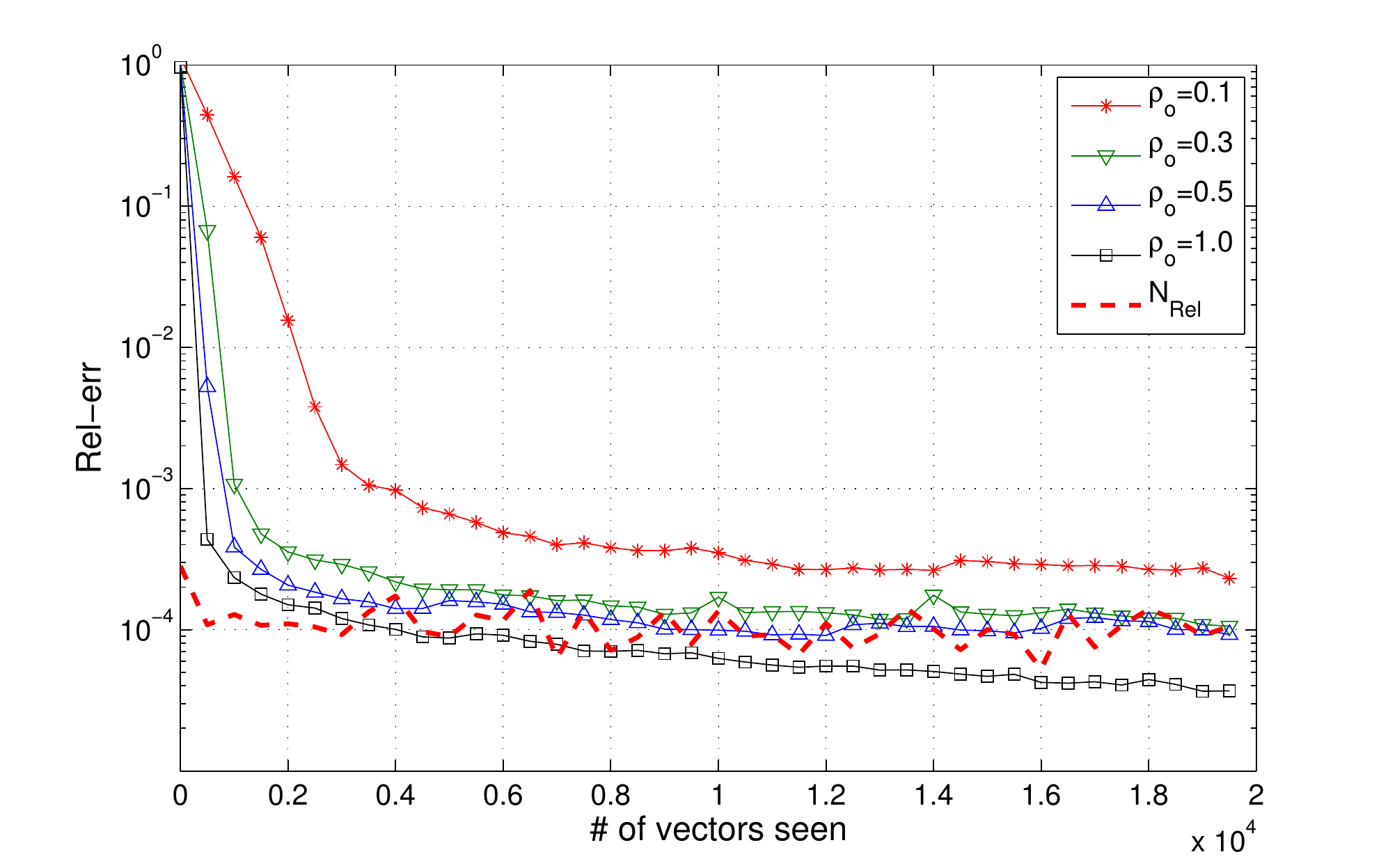} 
		\caption{The performance of stationary subspace identification within 10\% outliers using partial information. We show the results with different sub-sampling ratios, from using just 10\% information to using full information. The ambient dimension is $n=500$ and subspace dimension $d=5$, and all observed entries are also perturbed by small Gaussian noise with the variance of $\omega^2 = 10^{-5}$. }
		\label{fig:partial}
	\end{center}	
\end{figure}

In Figure~\ref{fig:partial}, we consider GRASTA's error performance for varying sub-sampling rates. Here the fraction of corrupted values is fixed at 10\%. We can see that again, even with a 30\% sampling rate, the relative error quickly reaches the relative noise power.

Now we wish to take a closer look at the case when we have both dense outlier corruption and subsampling of the signal. This is an important scenario for applications where the ``outlier corruption'' is a signal of interest obscuring a low-rank background signal, and we wish to subsample in order to improve computational complexity. For example this would apply to anomaly detection problems or to the problem of separation of background and moving objects in video as we show in Section~\ref{sec:realtimevideo}.

\begin{figure}[h!]
	\begin{center}
		\includegraphics[width=0.7\textwidth]{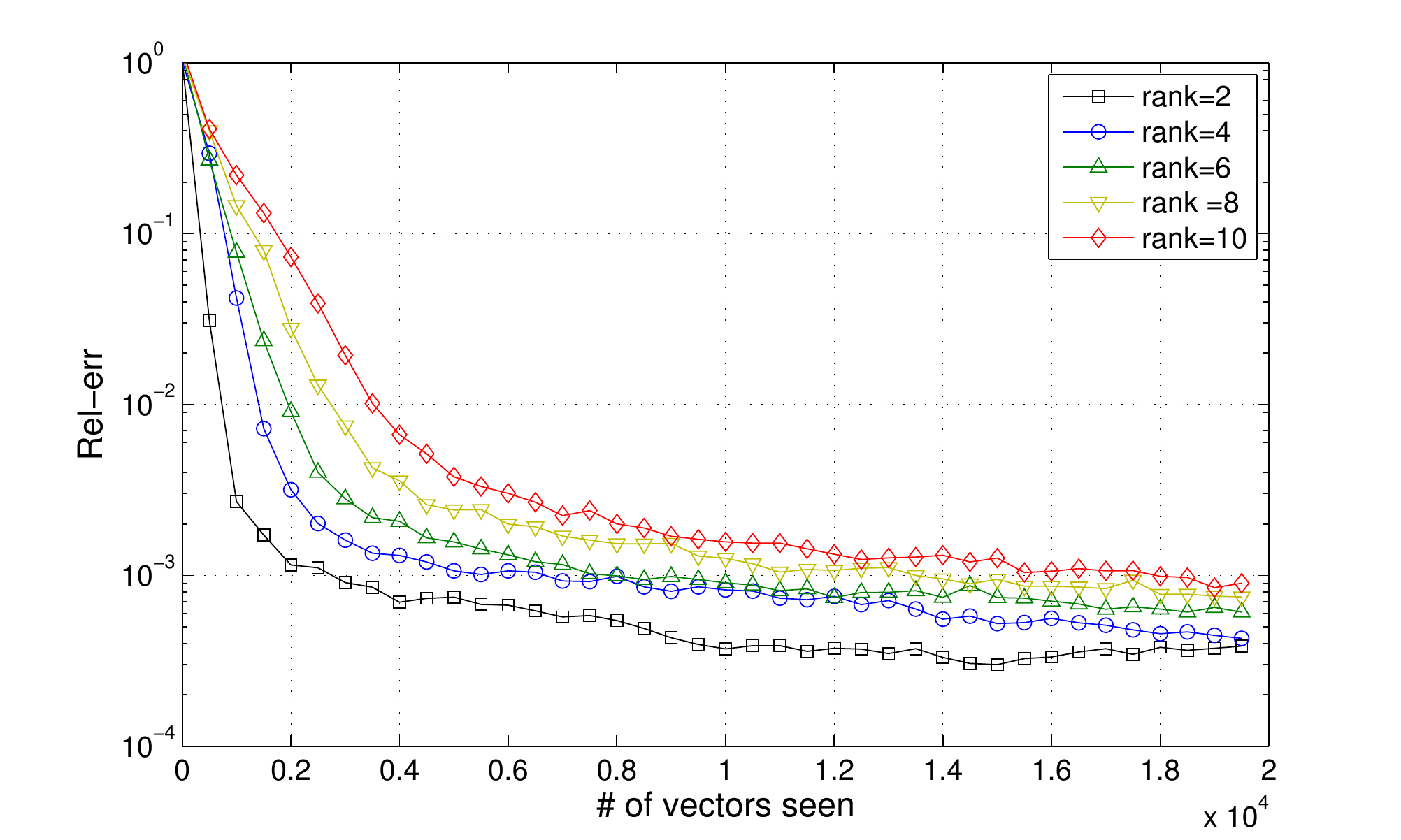} 
		\caption{The performance of subspace identification under dense error corruption. Here $\rho_s = 0.5$ which means 50\% entries of every data vector are corrupted, and we only observe 50\% entries of each vector. The ambient dimension is $n=500$, we vary the inherent dimension $d$, and all observed entries are also perturbed by small Gaussian noise with the variance of $10^{-5}$. We generate $20000$ vectors to show the performance over time.}
		\label{fig:dense_err}
	\end{center}	
\end{figure}

Figure~\ref{fig:dense_err} illustrates that even when the vector is highly corrupted with 50\% added outliers, GRASTA can identify the underlying low-rank subspace even with only 50\% of the entries. We vary the dimension (or rank) of the underlying subspace, and because of this there is not one relative noise power benchmark to compare against; however we see that the trend is similar to those in previous figures.

\subsection{Dynamic Subspace Tracking}

The fact that GRASTA operates one vector at a time allows it to track an evolving subspace. In this section we show GRASTA's performance under two models of evolving subspaces. In these experiments, we use the same set-up as before: $n=500$, $d=5$, and $v_t$ is generated by Equation~\eqref{eq:generative_model}, except that $U_{true} = U[t]$, i.e. the subspace we wish to estimate varies with time $t$:

\begin{equation}
	v_t = U[t] w_t + s_t + \zeta_t \;\;.
\end{equation}

\subsubsection{Rotating Subspace Tracking}
\begin{figure}[!h]
	\begin{center}
		\includegraphics[width=0.7\textwidth]{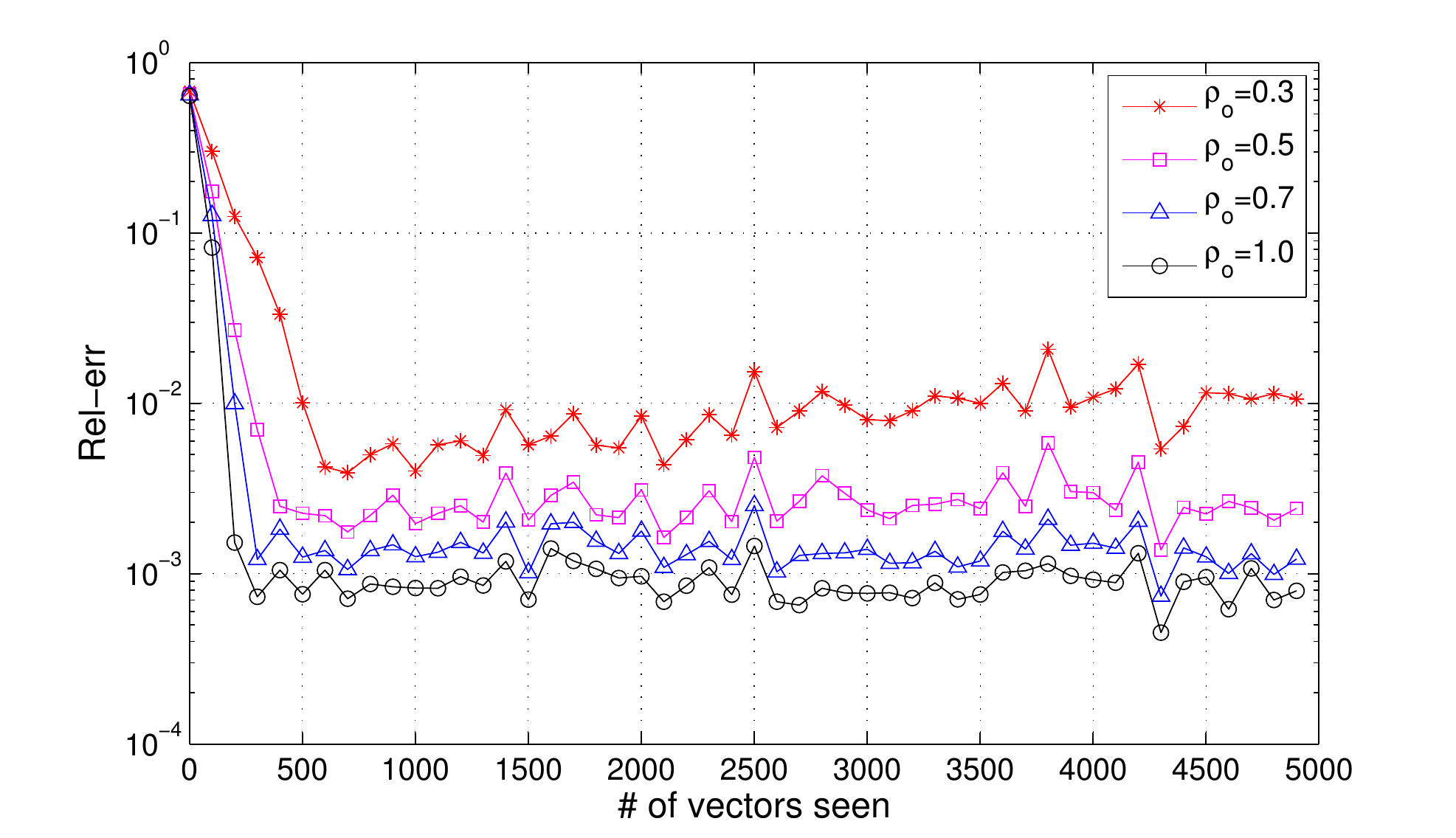} 
		\caption{The performance of tracking a rotating subspace within 10\% outliers. At every time the subspace rotates $\delta = 10^{-5}$. The noise variance is $\omega^2=10^{-5}$. We show  the results with varying sub-sampling. }
		\label{fig:fig_rotate}
	\end{center}	
\end{figure}

We use the following ordinary differential equation to simulate a rotating subspace:
\begin{equation}\label{eq:rotation}
	\dot{U}=BU, \hspace{.1in} U[0]=U_0
\end{equation}
where $B$ is a skew-symmetric  matrix. Consequently, the subspace $U[t]$ is updated via $$U[t] = e^{t\delta B}U_0\;,$$ where $\delta$ controls the amount of rotation of with each time step $t$. 
As we see in Figure~\ref{fig:fig_rotate}, for the rotation parameter $\delta$ fixed at $10^{-5}$, GRASTA successfully latches on and tracks the rotating subspace. 

\subsubsection{Sudden Subspace Change Tracking}

\begin{figure}[!h]
	\begin{center}
		\includegraphics[width=0.7\textwidth]{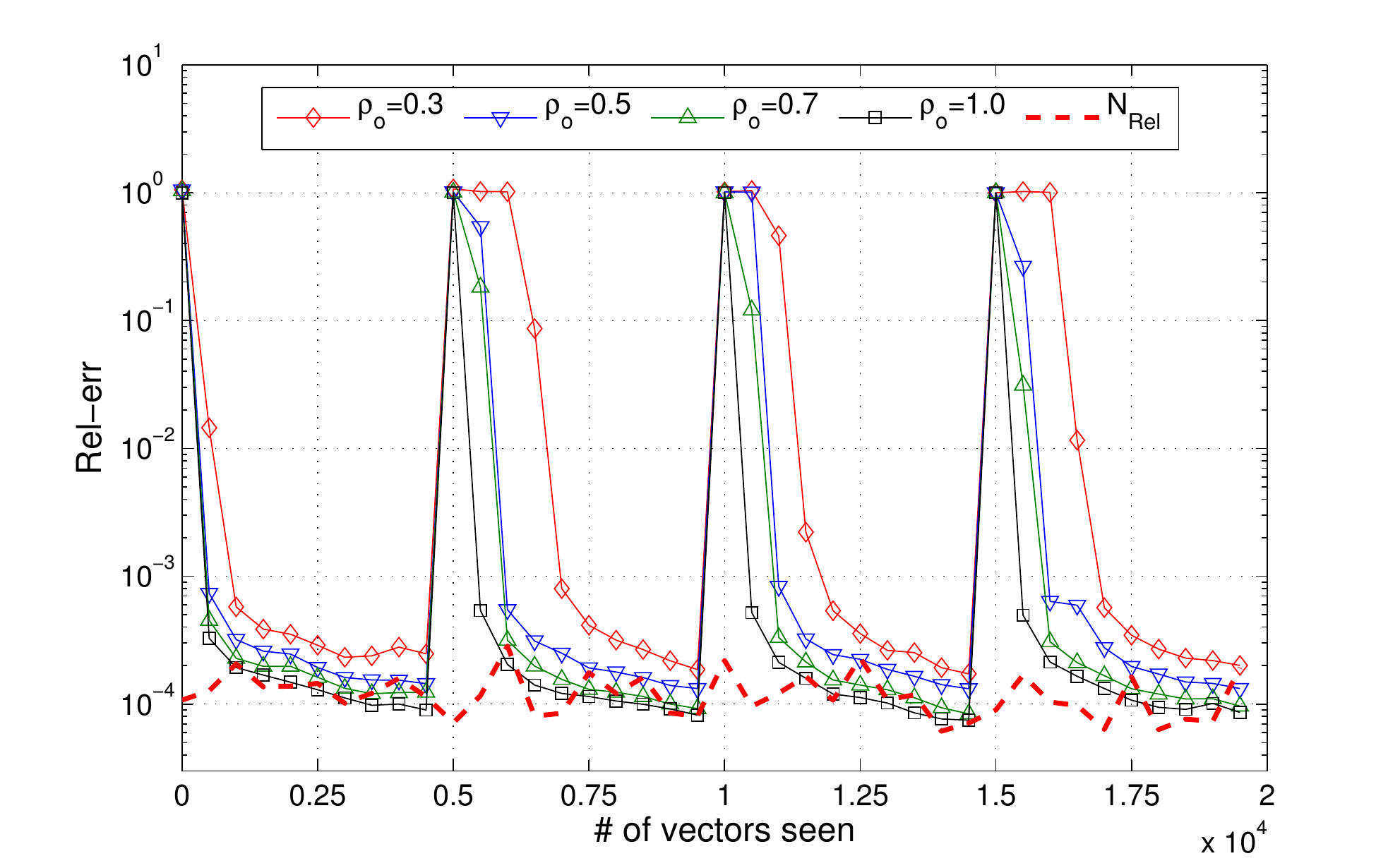} 
		\caption{The performance of subspace tracking within 10\% outliers. At time 5000, 10000, 15000, and 20000, the subspace undergoes a sudden change. Parameters are again $n=500$, $d=5$, $\omega^2=10^{-5}$. We show  the results with different sub-sampling ratios. }
		\label{fig:fig_sudden}
	\end{center}	
\end{figure}

For this experiment, we wanted to see the behavior of GRASTA when the subspace experienced a sudden dramatic change. At intervals of 5000 vectors, we randomly changed the true subspace to a new random subspace. 
The results are in Figure~\ref{fig:fig_sudden}. Again from these simulations we see that GRASTA successfully identifies the subspace change and tracks the subspace again.

\subsection{Comparison with Robust PCA}
Here we compare GRASTA with RPCA on the recovery of corrupted low-rank matrices.  
For RPCA we use~\cite{lin2010augmented}, or the IALM (Inexact Augmented Lagrange Multiplier) method\footnote{The code we used is available here: \url{http://perception.csl.uiuc.edu/matrix-rank/sample_code.html}. We downloaded it in April 2011.}.

The corrupted matrices can be written as $M = L + S + N$, where $L$ are the low-rank matrices we want to recover, $S$ are the sparse outlier matrices, and $N$ are the Gaussian noise matrices with small variance relative to the sparse outliers. We use $d=5$ matrices of size $2000 \times 2000$ to do the comparison. The low-rank matrices $L$ are generated by the same method of the previous robust matrix completion experiments: as the product of two $2000 \times 5$ factors $Y_L$ and $Y_R$ with i.i.d. Gaussian entries. The sparse outlier matrices $S$ are generated by selecting a fraction of entries uniformly at random without replacement, whose values are set according to Gaussian distribution with the maximum of $L$ as the variance\footnote{We note here that in~\cite{lin2010augmented}, the authors use a uniform distribution for the outliers, as opposed to Gaussian. The authors in~\cite{CandesRPCA09} use $\pm 1$ Bernoulli variables. Gaussian is the most challenging case, because more outliers will be near zero and confuse the estimation.}. We vary the fraction of corruptions from $10\%$ (sparse outliers) to $50\%$ (dense outliers), and we also vary the variance Gaussian noise matrices $N$ from a moderate perturbation of $\omega^2 = 10^{-4}$ to a larger perturbation of $\omega^2 = 10^{-3}$. For GRASTA we cycled through the matrix columns twice and used a maximum of $K=60$ iterations of the ADMM algorithm; we used a maximum of $20$ iterations of IALM. 

Table \ref{tbl:cmp_rpca_5} shows the results of the comparison. We ran RPCA with full data, GRASTA with full data, and then GRASTA with various levels of subsampling. When there are very few outliers and little noise, RPCA achieves a reasonable error rate at computational speeds similar to GRASTA. However with an increase in noise or fraction of outliers, GRASTA achieves good error performance in much less time. As a particular example, when $\omega^2 = 10^{-3}$ and the fraction of outliers is 30\%, in 69 seconds GRASTA with full data achieves better error performance than RPCA in 363 seconds, and GRASTA with 30\% subsampling achieves better error performance in only 23 seconds.

\begin{table}[!h]
\tiny
  \centering
  \begin{tabular}{|c|c|c|c|c|c|}
    \hline
    \multicolumn{2}{|c|}{} 						& GRASTA $\rho_o=1.0$ & GRASTA $\rho_o=0.5$& GRASTA $\rho_o=0.3$  & IALM \\
    \hline
     \multirow{3}{*}{$\rho_s=0.1$} 
     & $\omega^2=1 \times10^{-4}$  	& 1.38 E-4 / 56.62 sec &	 2.03 E-4 / 30.46 sec	&	2.73 E-4 / 20.51 sec		& 5.80 E-5 / 35.26 sec  \\
    \cline{2-6}
     & $\omega^2=5 \times 10^{-4}$  & 3.64 E-4 / 58.31 sec &	 4.65 E-4 / 31.23 sec	&	6.07 E-4 / 20.79 sec		& 1.67 E-3 / 93.16 sec  \\
    \cline{2-6}
     & $\omega^2=1 \times 10^{-3}$  & 7.64 E-4 / 59.55 sec & 9.59 E-4 / 31.81 sec	&	1.23 E-3 / 20.66 sec		& 3.64 E-3 / 117.76 sec  \\
    \hline
     \multirow{3}{*}{$\rho_s=0.3$} 
     & $\omega^2=1 \times10^{-4}$  	& 4.65 E-4 / 67.90 sec & 7.28 E-4 / 35.10 sec	&	1.06 E-3 / 22.96 sec		& 1.80 E-4 / 232.26 sec  \\
    \cline{2-6}
     & $\omega^2=5 \times 10^{-4}$  & 6.13 E-4 / 67.19 sec &	 9.08 E-4 / 34.53 sec	&	1.26 E-3 / 22.63 sec		& 2.64 E-3 / 324.26 sec  \\
    \cline{2-6}
     & $\omega^2=1 \times 10^{-3}$  & 9.87 E-4 / 69.06 sec &	 1.44 E-3 / 35.61 sec	&	1.93 E-3 / 22.85 sec		& 5.62 E-3 / 362.62 sec  \\
    \hline
     \multirow{3}{*}{$\rho_s=0.5$} 
     & $\omega^2=1 \times10^{-4}$  	& 1.26 E-3 / 83.11 sec & 2.05 E-3 / 39.90 sec	&	3.58 E-3 / 25.33 sec		& 1.43 E-1 / 341.01 sec  \\
    \cline{2-6}
     & $\omega^2=5 \times 10^{-4}$  & 1.33 E-3 / 81.51 sec &	 2.24 E-3 / 40.33 sec	&	3.93 E-3 / 25.38 sec		& 1.45 E-1 / 351.26 sec  \\
    \cline{2-6}
     & $\omega^2=1 \times 10^{-3}$  & 1.64 E-3 / 82.23 sec & 2.85 E-3 / 41.91 sec	&	5.08 E-3 / 25.67 sec		& 1.62 E-1 / 372.21 sec  \\
    \hline    
    
  \end{tabular}
	\caption{Recovery of corrupted low-rank matrices; a comparison between GRASTA and Robust PCA. We use full information of the corrupted matrices to do robust PCA, and vary the sub-sampling rate $\rho_o$ from $0.3$ to $1.0$ (30\% of the data to full information), to perform GRASTA. The matrix is $2000 \times 2000$, rank=$5$. We vary the fraction of corruptions from sparse outliers 10\% to dense outliers 50\%, and also vary the Gaussian noise variance $\omega^2$  from moderate noise perturbation $\omega^2=1\times 10^{-4}$ to relative strong noise corruption $\omega^2=1\times 10^{-3}$. }
	\label{tbl:cmp_rpca_5}
\end{table}

\subsection{Realtime Video Background Tracking and Foreground Detection }
\label{sec:realtimevideo}

In this subsection we discuss the application of GRASTA to the prominent problem of realtime separation of foreground objects from the background in video surveillance. Imagine we had a video with only the background: When the columns of a single frame of this video are stacked into a single column, several frames together will lie in a low-dimensional subspace. In fact if the background is completely static, the subspace would be one-dimensional. That subspace can be estimated in order to identify and separate the foreground objects; if the background is dynamic, subspace tracking is necessary. GRASTA is uniquely suited for this burgeoning application.

Here we consider three scenarios in the video tasks, with a spectrum of challenges for subspace tracking. In the first we have a video with a static background and objects moving in the foreground. In the second, we have a video with a still background but with changing lighting. In the third, we simulate a panning camera to examine GRASTA's performance with a dynamic background. The results are summarized in Table~\ref{tbl:realtime}.

\subsubsection{Static Background}
If the video background is known to be static or near static, we can use GRASTA to track the background and separate the moving foreground objects in real-time. Since the background is static, we use GRASTA first to identify the background, and then we use only Algorithm~\ref{alg:SRP} to separate the foreground from the background. More precisely we do the following:  

\begin{enumerate}
	\item
	Randomly select a few frames of the video to train the static low-rank subspace $U$. In our experiments, we select frames randomly from the entire video; however for real-time processing these frames may be chosen from initial piece of the video, as long as we can be confident that every pixel of the background is visible in one of the selected frames. The low-rank subspace $U$ is then identified from these frames using partial information. We use $30\%$ of the pixels, select $50$ frames for training, and set RANK = 5 in all the following experiments.	
	
	\item 
	Once the video background $BG$  has been identified as a subspace $U$, separating the foreground objects $FG$  from each frame can be simply done using Equation \eqref{eq:video_separation}, where the weight vector $w_t$ can be solved for via Algorithm \ref{alg:SRP}, again from a small subsample of each frame's pixels. 
	
\begin{equation}\label{eq:video_separation}
\left\{ \begin{array}{l}
	BG = Uw_t \\
	FG =  video(t) - BG
\end{array} \right.
\end{equation}

\end{enumerate}

\begin{figure}[!h]
	\begin{center}
		\includegraphics[width=0.6\textwidth]{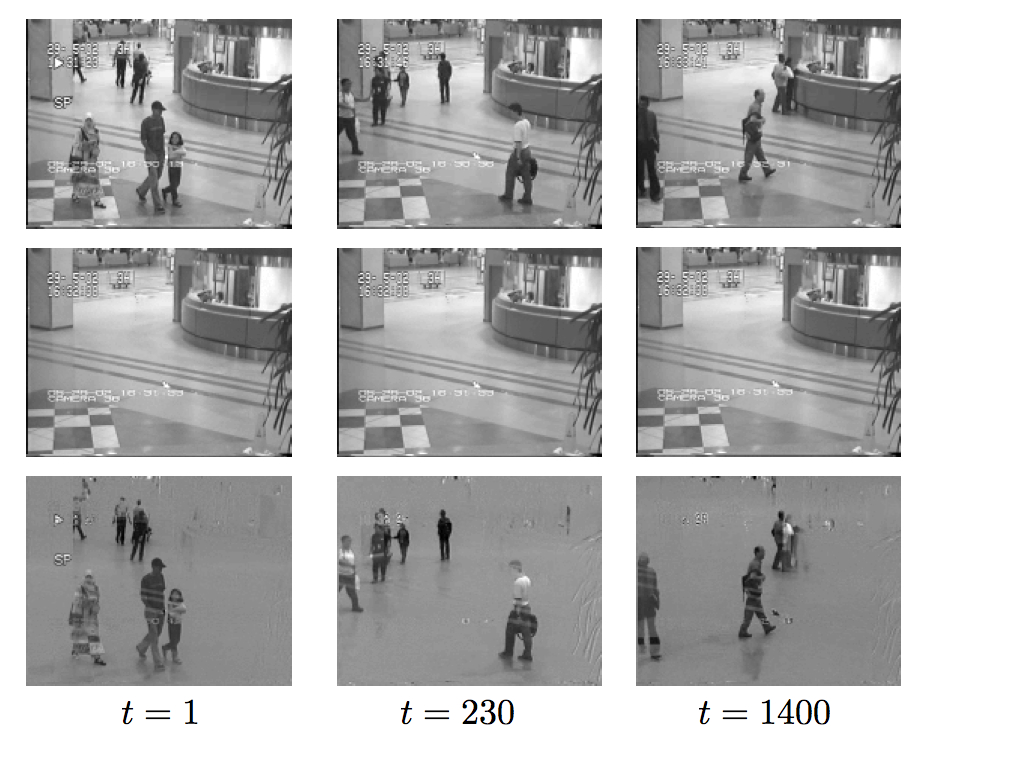} 
		\caption{Real-time video background and foreground separation from partial information. We show the separation quality at $t=1, 230, 1400$. The resolution of the video is $144 \times 176$. The first row is the original video frame at each time; the middle row is the recovered background at each time only from $5\%$ information; and bottom row is the foreground calculated by Equation \eqref{eq:video_separation}.}
		\label{fig:fig_static_hall}
	\end{center}	
\end{figure}

\begin{figure}[!h]
	\begin{center}
		\includegraphics[width=0.6\textwidth]{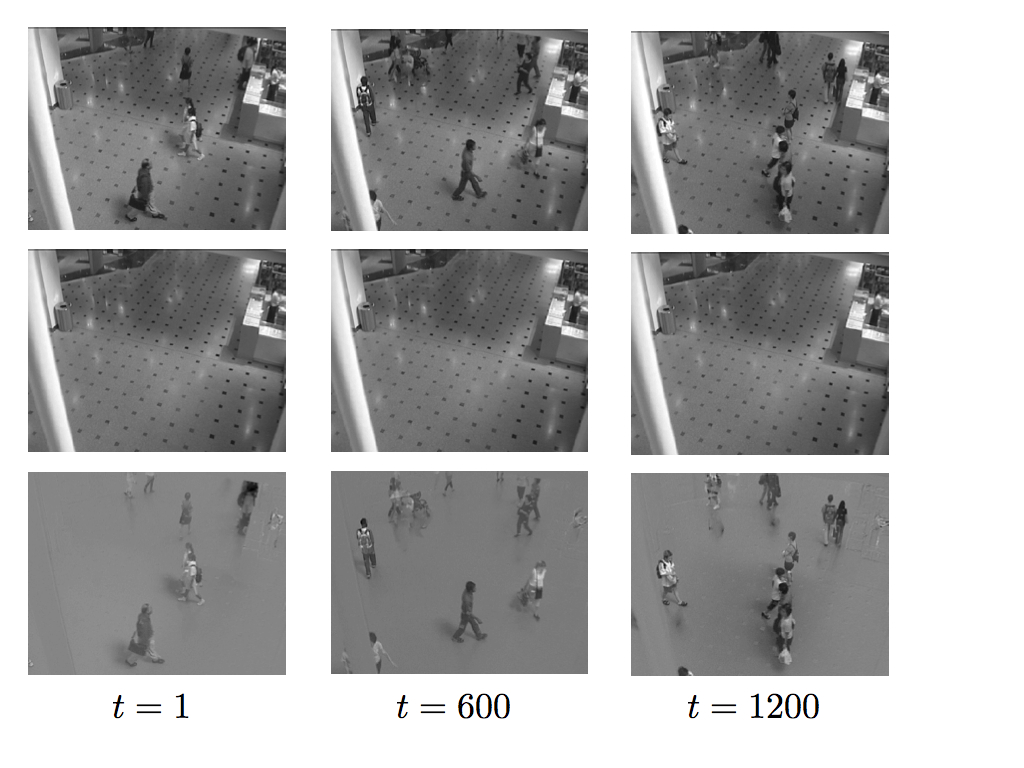} 
		\caption{Real-time video background and foreground separation from partial information. We show the separation quality at $t=1, 600, 1200$. The resolution of the video is $320 \times 256$. The first row is the original video frame at each time; the middle row is the recovered background at each time only from $1\%$ information; and bottom row is the foreground calculated  by Equation \eqref{eq:video_separation}.}
		\label{fig:fig_static_shopping}
	\end{center}	
\end{figure}

Table \ref{tbl:realtime} shows the real-time\footnote{We comment here that to call something ``real-time'' processing of course will depend on one's application requirements and hardware (camera frame capture rate, in the example of video processing). For example, standard 35mm film video uses 24 unique frames per second. The maximum frame rate for most CCTVs is 30 frames per second.} video separation results. From the first experiment, we use the ``Hall" dataset from~\cite{Li04} which consists of $3584$ frames each with resolution $144 \times 176$. We let GRASTA cycle $5$ times over the $50$ training frames just from $30\%$ random entries of each frame to get the stationary subspace $U$. Training the subspace costs $6.9$ seconds. Then we perform background and foreground separation for all frames in a streaming fashion, and when dealing with each frame we only randomly observe $5\%$ entries. The separation task is performed by Equation \eqref{eq:video_separation}, and the separating time is $62.5$ seconds, which means we achieve $57.3$ FPS (frames per second) real-time performance. Figure \ref{fig:fig_static_hall} shows the separation quality at $t=1, 230, 1400$. In order to show GRASTA can handle higher resolution video effectively, we use the ``Shopping Mall"~\cite{Li04} video with resolution $320 \times 256$ as the second experiment. We also do the subspace training stage with the same parameter settings as ``Hall". We do the background and foreground separation only from $1\%$ entries of each frame. For ``Shopping Mall'' the separating time is $39.1$ seconds for total $1286$ frames. Thus we achieve $32.9$ FPS real-time performance. Figure \ref{fig:fig_static_shopping} shows the separation quality at $t=1, 600, 1200$. In all of these video experiments we used a maximum of $K=20$ iterations of the ADMM algorithm per subspace update. The details of each tracking set-up are described in Table~\ref{tbl:video_description}.

\begin{table}[!h]
\tiny
\begin{center}
	\begin{tabular}{ c | c | c | c | c | c }
	\hline
		Dataset &  Resolution & Total Frames &  Training Time & Tracking and & FPS\\
		& & &  & Separating Time& \\		\hline
		Hall & 144$\times$176 & 3584	& 6.9 sec & 62.5 sec & 57.3 \\
		Shopping Mall & 320$\times$256 & 1286	 & 23.2 sec & 39.1 sec & 32.9 \\			
		Lobby &  144$\times$176 & 1546  & 3.9 sec & 71.3 sec & 21.7	\\ 
		Hall with Virtual Pan (1) & 144$\times$88 & 3584	 & 3.8 sec& 191.3 sec& 18.7 \\ 
		Hall with Virtual Pan (2) & 144$\times$88 & 3584 & 3.7 sec & 144.8 sec & 24.8 \\  \hline
	\end{tabular}
\end{center}
	\caption{Real-time video background and foreground separation by GRASTA. Here we use three different resolution video datasets, the first two with static background and the last three with dynamic background. We train from 50 frames; in the first two experiments they are chosen randomly, and in the last three they are the first 50 frames. In all experiments, the subspace RANK = 5. }
		\label{tbl:realtime}
\end{table}

\begin{table}[!h]
\tiny
\begin{center}
	\begin{tabular}{ c | c | c | c | c | c }
	\hline
		Dataset &  Training & Tracking & Separation & Training Algorithm & Tracking/Separation Algorithm\\ 
		& Sub-Sampling & Sub-Sampling & Sub-Sampling & & \\ \hline
		Hall & 30\% & - & 5\% & Full GRASTA Alg~\ref{alg:GLAD} & Alg~\ref{alg:SRP}+Eqn~\ref{eq:video_separation} \\
		Shopping Mall & 30\% & - & 1\% & Full GRASTA Alg~\ref{alg:GLAD} & Alg~\ref{alg:SRP}+Eqn~\ref{eq:video_separation}  \\
		Lobby & 30\% & 30\% & 100\% &  Full GRASTA Alg~\ref{alg:GLAD} &  Full GRASTA Alg~\ref{alg:GLAD}	\\ 
		Hall with Virtual Pan (1) & 100\% & 100\% & 100\%& Full GRASTA Alg~\ref{alg:GLAD} &  Full GRASTA Alg~\ref{alg:GLAD} \\ 
		Hall with Virtual Pan (2) & 50\% & 50\% & 100\%& Full GRASTA Alg~\ref{alg:GLAD} &  Full GRASTA Alg~\ref{alg:GLAD} \\\hline
	\end{tabular}
\end{center}
	\caption{Here we summarize the approach for the various video experiments. When the background is dynamic, we use the full GRASTA for tracking. We used $K=20$ iterations of the ADMM algorithm for all video experiments.}
		\label{tbl:video_description}
\end{table}

\subsubsection{Dynamic Background: Changing Lighting}

\begin{figure}[!h]
	\begin{center}
		\includegraphics[width=.9\textwidth]{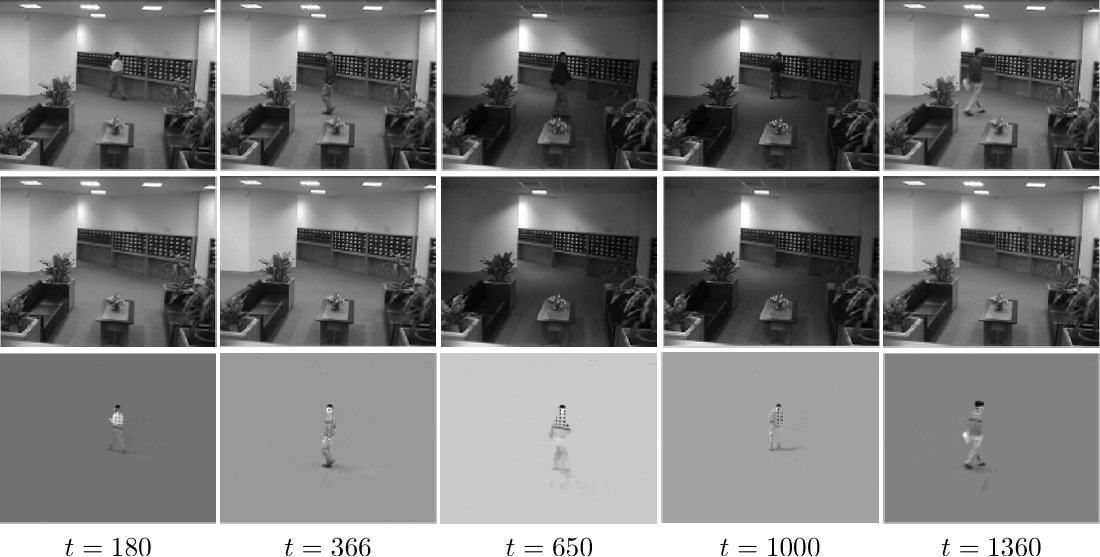} 
		\caption{Real-time video background and foreground separation from partial information. We show the separation quality at $t=180, 366, 650, 1000, 1360$. The resolution of the video is $144 \times 176$ and has a total of 1546 frames. The first row is the original video frame at each time; the middle row is the recovered background at each time only from $30\%$ information; and bottom row is the foreground calculated by Algorithm~\ref{alg:SRP} using full information. The differing background colors of the bottom row is simply an artifact of colormap in Matlab.}
		\label{fig:fig_lobby}
	\end{center}	
\end{figure}

Here we want to consider a problem where the lighting in the video is changing throughout. We use the ``Lobby" dataset from~\cite{Li04}, which has $1546$ frames, each $144 \times 176$ pixels. In order to adjust to the lighting changes, GRASTA tracks the subspace throughout the video; that is, unlike the last two experiments, we run the full GRASTA Algorithm~\ref{alg:GLAD} for every frame. We use 30\% of the pixels of every frame to do this update and 100\% of the pixels to do the separation. Again, see the numerical results in Table~\ref{tbl:realtime}. The results are illustrated in Figure~\ref{fig:fig_lobby}.

\subsubsection{Dynamic Background: Virtual Pan}
\begin{figure}[!h]
	\begin{center}
		\includegraphics[width=0.5\textwidth]{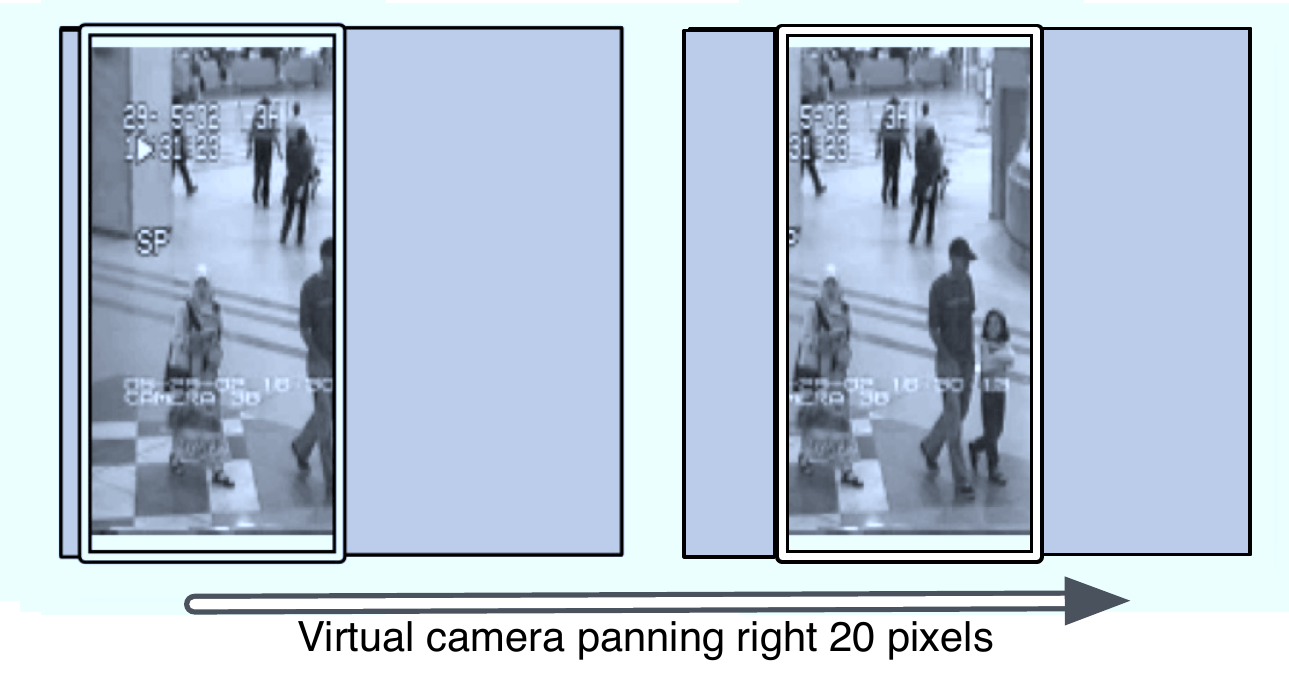} 
		\caption{ Demonstration of panning the "virtual camera" right $20$ pixels.}
		\label{fig:fig_virual_camera}
	\end{center}	
\end{figure}

In the last experiment, we demonstrate that GRASTA can effectively track the right subspace in video with a dynamic background. We consider panning a "virtual camera" from left to right and right to left through the video to simulate a dynamic background. Periodically, the virtual camera pans 20 pixels.
The idea of the virtual camera  is illustrated cleanly with Figure \ref{fig:fig_virual_camera}. 

\begin{figure}[!h]
	\begin{center}
		\includegraphics[width=1.0\textwidth]{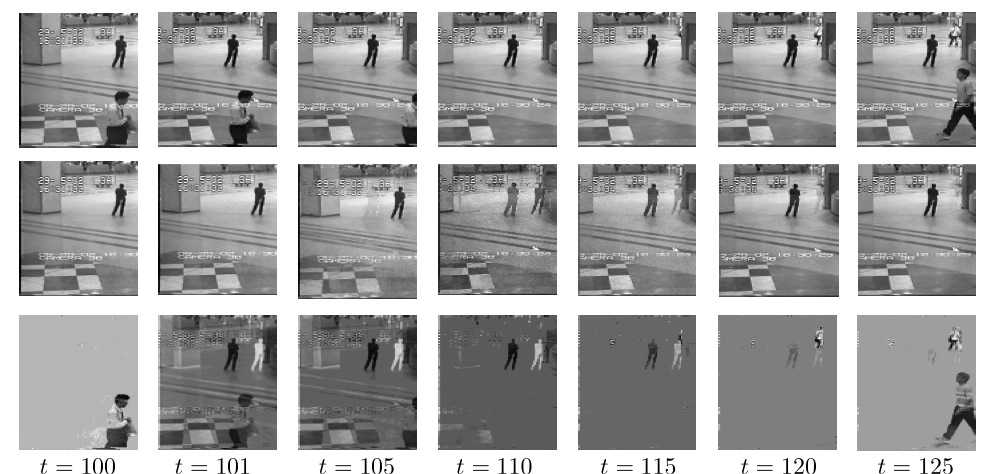} 
		\caption{Real-time dynamic background tracking and foreground separation. At time $t=101$, the virtual camera slightly pans to right $20$ pixels. We show how GRASTA quickly adapts to the new subspace at $t=100, 105, \ldots, 125$. The first row is the original video frame at each time; the middle row is the tracked background at each time; the bottom row is the separated foreground at each time.}
		\label{fig:fig_dynamic}
	\end{center}	
\end{figure}

We choose ``Hall" as the original dataset. The original resolution is $144 \times 176$, and we set the scope of the virtual camera to have the same height but half the width, so the resolution of the virtual camera is $144 \times 88$. We set the subspace $RANK=5$. Figure \ref{fig:fig_dynamic} shows how GRASTA can quickly adapt to the changed background in just $25$ frames when the virtual camera pans $20$ pixels to the right at $t=101$.  We also let GRASTA track all the $3584$ frames and do the separation task for all frames. When we use 100\% of the pixels for the tracking and separation, the total computation time is $191.3$ seconds, or $18.7$ FPS, and adjusting to a new camera position after the camera pans takes $25$ frames as can be seen in Figure~\ref{fig:fig_dynamic}. When we use 50\% of the pixels for tracking and 100\% of the pixels for separation, the total computation time is $144.8$ seconds or $24.8$ FPS, and the adjustment to the new camera position takes around 50 frames.

\section{Discussion and Future Work}
\label{sec:conclusion}

In this paper we have presented a robust online subspace tracking algorithm, GRASTA. The algorithm estimates a low-rank model from noisy, corrupted, and incomplete data, even when the best low-rank model may be changing over time.

Though this work presents some very successful algorithms, many questions remain. First and foremost, because the cost function in Equation~\eqref{eq:L1_distance} has the subspace variable $U$ which is constrained to a non-convex manifold, the resulting optimization is non-convex. A proof of convergence to the global minimum of this algorithm is of great interest.

GRASTA uses alternating minimization, alternating first to estimate $(s,w,y)$ and then fixing this triple of variables to estimate $U$. Observe that if $(s,w,y)$ are correct estimates, we could then estimate $U$ {\em without} the robust cost function. This would be quite useful in situations when speed is of utmost importance, as the GROUSE subspace update is faster than the GRASTA subspace update. Of course, knowing when $(s,w,y)$ are accurate is a very tricky business. Exploring this tradeoff is part of our future work.

We have shown that one of the very promising applications of GRASTA is that of separating background and foreground in video surveillance. We are very interested to apply GRASTA to more videos with dynamic backgrounds: for example, natural background scenery which may blow in the wind. In doing this we will study the resulting trade-off between the kinds of movement that would be captured as part of the background and the movement that would be identified as foreground.

\section*{Acknowledgments}

The authors would like to thank IPAM, the Institute for Pure and Applied Mathematics, and the Internet Multi-Resolution Analysis program, which brought them together to work on this problem. We also thank Rob Nowak and Ben Recht for their thoughtful suggestions.

\small 
\bibliographystyle{plain}
\bibliography{GRASTA2011}

%\input{biography}

\end{document}